\documentclass[%
reprint,
twocolumn,
superscriptaddress,
amsmath,
amssymb,
aps,
prl,
]{revtex4-1}

\usepackage{graphicx}
\usepackage{dcolumn}
\usepackage{bm}
\usepackage{hyperref}
\usepackage{amsmath}
\usepackage{amsfonts}
\usepackage{amssymb}
\usepackage{braket}
\usepackage{bbold}
\usepackage{tabularx}
\usepackage{booktabs}
\newcolumntype{Y}{>{\centering\arraybackslash}X}
\usepackage{multirow}
\usepackage{float}

\mathchardef\mhyphen="2D
\begin{document}
\title{Infinite-layer nickelates as Ni-$e_g$ Hund's metals}
\date{\today}

\author{Byungkyun Kang}
\affiliation{
  Condensed Matter Physics and Materials Science Department, 
  Brookhaven National Laboratory, Upton, NY 11973, USA
}
\author{Corey Melnick}
\affiliation{
  Condensed Matter Physics and Materials Science Department, 
  Brookhaven National Laboratory, Upton, NY 11973, USA
}
\author{Patrick Semon}
\affiliation{
  Condensed Matter Physics and Materials Science Department, 
  Brookhaven National Laboratory, Upton, NY 11973, USA
}
 \author{Siheon Ryee}
 \affiliation{Department of Physics, KAIST, Daejeon 34141, Republic of Korea}
 \author{Myung Joon Han}
 \affiliation{Department of Physics, KAIST, Daejeon 34141, Republic of Korea}
 
\author{Gabriel Kotliar}
\affiliation{
  Department of Physics and Astronomy, Rutgers University, New Jersey 08854, USA
}
\affiliation{
  Condensed Matter Physics and Materials Science Department, 
  Brookhaven National Laboratory, Upton, NY 11973, USA
}

\author{Sangkook Choi}
\email[E-mail me at: ]{sachoi@bnl.gov}
\affiliation{
  Condensed Matter Physics and Materials Science Department, 
  Brookhaven National Laboratory, Upton, NY 11973, USA
}

\begin{abstract}

  The recent and exciting discovery of superconductivity in the hole-doped infinite-layer nickelate Nd$_{1-\delta}$Sr$_\delta$NiO$_2$ draws strong attention to correlated quantum materials. From a theoretical view point, this new class of unconventional superconducting materials provides an opportunity to unveil new physics in correlated quantum materials. Here we study the temperature and doping dependence of the local spectrum as well as the charge, spin and orbital susceptibilities \textit{from first principles}. By using \textit{ab initio} LQSGW+DMFT methodology, we show that onsite Hund's coupling in Ni-$d$ orbitals gives rise to multiple signatures of Hund's metallic phase in Ni-$e_g$ orbitals. The proposed picture of the nickelates as an $e_g$ (two orbital) Hund's metal differs from the picture of the Fe-based superconductors as a five orbital Hund's metal as well as the picture of the cuprates as doped charge transfer insulators. Our finding unveils a new class of the Hund’s metals and has potential implications for the broad range of correlated two orbital systems away from half-filling. 

\end{abstract}
\maketitle
\textit{Introduction}. Although the mechanisms of unconventional superconductivity remain elusive, the discoveries of new classes of unconventional superconductors have proliferated experimentally. These experimental efforts revived the interest in correlated quantum materials and provided opportunities to unveil new physics hidden within them. To illustrate, in the cuprate superconductors \cite{bednorz_possible_1986}, superconductivity emerges from the bad metallic states realized by doping a charge transfer insulator \cite{zaanen_band_1985}. Strong electron correlation in the bad metallic normal states arises due to the proximity to Mott insulator transition  \cite{mott_discussion_1937,mott_basis_1949}, i.e., ``Mottness''. According to the theory of conventional superconductors, it is improbable that this bad-metallic phase would support superconductivity. This motivated the theoretical proposals of superconducting pairing mechanisms beyond the Bardeen-Cooper-Schrieffer (BCS) paradigm \cite{lee_doping_2006,garg_strong_2008,keimer_quantum_2015}.  This in turn lead to the discovery of other unconventional superconductors wherein a superconducting phase emerged from the bad-metal "parent" state in a different way. For example, in the multi-orbital Fe-based superconductors \cite{kamihara_iron-based_2006,kamihara_iron-based_2008}, the on-site Hund’s coupling ($J$) promotes bad metallic behavior in their normal phase \cite{de_medici_janus-faced_2011,de_medici_hunds_2011,georges_strong_2013,isidori_charge_2019}. This gives rise to the new concept of ``Hundness''. Hundness-induced correlated metals, Hund's metals \cite{haule_coherenceincoherence_2009,yin_kinetic_2011}, play the role of a reliable reference system for Fe-based superconducting materials \cite{georges_strong_2013,haule_coherenceincoherence_2009,yin_kinetic_2011,de_medici_hunds_2017,lanata_orbital_2013,villar_arribi_hund-enhanced_2018,bascones_orbital_2012,ryee_nonlocal_2020} and ruthenates \cite{georges_strong_2013,werner_spin_2008,mravlje_coherence-incoherence_2011,hoshino_superconductivity_2015,mravlje_thermopower_2016}.

Recently, the thrilling discovery of Ni-based superconductors \cite{li_superconductivity_2019,li_superconducting_2020,gu_two_2020,zeng_phase_2020} turns the spotlight on correlated quantum materials and their unconventional superconductivity \cite{sawatzky_superconductivity_2019,xiang_magnetic_2020}. NdNiO$_2$ and infinite-layer cuprates, e.g. CaCuO2, are isostructural \cite{hayward_sodium_1999,hayward_synthesis_2003}, where the two dimensional Ni-O plane is geometrically analogous to the Cu-O plane in the cuprate. The Ni-$d_{x2-y2}$ orbital of each Ni$^{1+}$ ion can be expected to be half-filled with an effective spin-1/2 on each site according to the oxidation state rules. In combination, this makes NdNiO$_2$ a promising cuprate analog \cite{botana_similarities_2020,kitatani_nickelate_2020,hirsch_hole_2019,wu_robust_2020,karp_many-body_2020,lang_where_2020}. 

However, the differences from cuprates are striking. Its parent compound is seldom regarded as a charge transfer insulator \cite{ikeda_improved_2013,ikeda_direct_2016,fu_electronic_2019,kitatani_nickelate_2020} and there is no sign of long-range magnetic orders \cite{hayward_synthesis_2003} down to 1.7 K. In addition, its parent compound shows a resistivity upturn upon cooling \cite{li_superconductivity_2019}, which is common in heavy-fermion superconductors and is often due to Kondo effects \cite{hepting_electronic_2020,lechermann_multiorbital_2020}. The sign change of the Hall coefficient implies that electrons as well as holes may play an important role in the materials properties \cite{li_superconductivity_2019}, implying its multi-orbital nature \cite{adhikary_orbital_2020,lechermann_multiorbital_2020,goodge_doping_2020}. Moreover, it is debatable whether the doped hole forms a spin singlet or triplet doublon with the original hole on a Ni ion \cite{jiang_critical_2020,zhang_type-ii_2020,hu_two-band_2019,werner_nickelate_2020,zhang_self-doped_2020,chang_hund-heisenberg_2019,sakakibara_model_2019}, suggesting possible Hund metal physics \cite{wang_hunds_2020,lechermann_multiorbital_2020,petocchi_normal_2020,lechermann_multiorbital_2020}. These similarities and differences to various unconventional superconductors are puzzling, but they do provide a chance to explore hidden aspects of electron correlation.

In this paper, we will investigate a new aspect of the multi-orbital physics in infinite-layer nickelates \textit{from first principles}. By using \textit{ab initio} linearized quasiparticle self-consistent GW (LQSGW) and dynamical mean-field theory (DMFT) method \cite{tomczak_qsgw_2015,choi_first-principles_2016,choi_comdmft:_2019}, we investigated the origin of the electron correlation in the infinite-layer nickelate normal phases. \textit{Ab initio} LQSGW+DMFT is a diagrammatically motivated \textit{ab initio} approaches for correlated electron systems. As a simplification of diagrammatically controlled full GW+EDMFT approach \cite{sun_extended_2002,biermann_first-principles_2003,nilsson_multitier_2017}, it calculates electronic structure by using \textit{ab initio} linearized quasiparticle self-consistent GW approaches \cite{kutepov_electronic_2012,kutepov_linearized_2017}. Then it adds one-shot correction to local electron self-energy by summing over all possible local Feynmann diagrams within DMFT \cite{georges_dynamical_1996,metzner_correlated_1989,muller-hartmann_correlated_1989,brandt_thermodynamics_1989,janis_new_1991,georges_hubbard_1992,jarrell_hubbard_1992,rozenberg_mott-hubbard_1992,georges_numerical_1992}. For the impurity orbital in the DMFT step, we choose a very localized orbital spanning a large energy window, which contains most strongly hybridized bands along with upper and lower Hubbard bands. Having chosen the shape of the correlated orbitals, all the other parameters to define DMFT problem are determined accordingly: double-counting energy within local GW approximation and Coulomb interaction tensor within constrained random phase approximation (cRPA) \cite{aryasetiawan_frequency-dependent_2004}. This method has been validated against various classes of correlated electron systems including paramagnetic Mott insulators La$_2$CuO$_4$\cite{choi_first-principles_2016}, Hund metal FeSe \cite{choi_comdmft:_2019}, and correlated narrow-gap semiconductors FeSb$_2$ \cite{chikina_correlated_2020}.

Within \textit{ab initio} LQSGW+DMFT, we found multiple signatures of Hundness associated with the Ni-$d$ subshell in the compounds. This finding differentiates the infinite-layer nickelates from the cuprates. In particular, we found that Hundness becomes apparent among the Ni-$e_g$ orbitals but not the Ni-$t_{2g}$ orbitals. This is a distinctive feature of the infinite-layer nickelates from Fe-based superconductors as five-orbital Hund's metals. 


\begin{figure}[t]
  \centering
  \includegraphics[width=1.0\columnwidth]{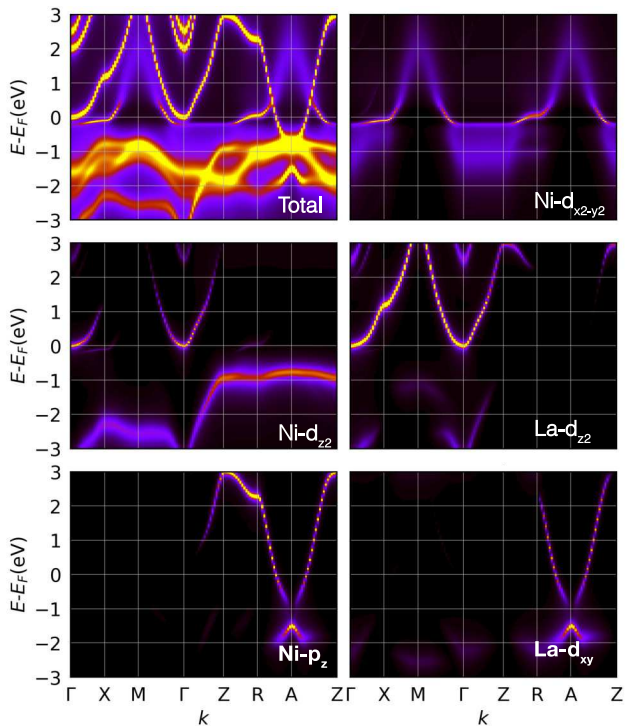}
  \caption{Total and orbital-resolved spectral function of La$_{0.8}$Ba$_{0.2}$NiO$_2$ along a high-symmetry line as calculated within \textit{ab initio} LQSGW+DMFT at T=300K. Of the two bands crossing the Fermi level, the lower energy band shows Ni-$d_{x^2-y^2}$ character, and the other, self-doping band at higher energy is a mixture of La-$d_{z^2}$, Ni-$d_{z^2}$, La-$d_{xy}$ and Ni-$p_{z}$. }
  \label{fig_spectral}
\end{figure}

\textit{Methods}.
Following the literature \cite{leonov_lifshitz_2020,ryee_induced_2020,botana_similarities_2020,wang_hunds_2020,olevano_ab_2020,bernardini_magnetic_2019,sakakibara_model_2019,hepting_electronic_2020}, we studied La$_{1-\delta}$Ba$_\delta$NiO$_2$ instead of Nd$_{1-\delta}$Sr$_{\delta}$NiO$_2$ to avoid the difficulty in the treatment of the Nd-$4f$ band. This is acceptable, as it has been reported that LaNiO$_2$ at the lattice parameters of NdNiO$_2$ has a similar electronic structure of NdNiO$_2$ within open Nd-$f$ core approximation \cite{botana_similarities_2020}. It has been experimentally demonstrated that the Nd-$4f$ states of Nd-based infinite layer nickelates are well localized and do not influence the relevant physics close to the Fermi level \cite{hepting_electronic_2020,rossi_orbital_2020}. The effect of Ba doping has been treated within the virtual crystal approximation. For its justification, please see the supplementary materials. For the LQSGW+DMFT scheme, the code ComDMFT \cite{choi_comdmft:_2019} was used. For the LQSGW part of the LQSGW+DMFT scheme, the code FlapwMBPT \cite{kutepov_linearized_2017} was used. For the details of electronic structure calculation, please see the supplementary materials. 

\textit{Results and Discussions}. The low-energy electronic structure of La$_{1-\delta}$Ba$_{\delta}$NiO$_2$ shows multi-orbital characters. In particular, the two bands crossing the Fermi-energy have substantial Ni-$e_g$ orbital character. Fig. \ref{fig_spectral} shows the electronic structure of La$_{0.8}$Ba$_{0.2}$NiO$_2$ within \textit{ab initio} LQSGW+DMFT. Consistent with the results obtained with other electronic structure methodologies such as DFT \cite{adhikary_orbital_2020,bernardini_magnetic_2019,botana_similarities_2020,jiang_electronic_2019,lechermann_late_2020,lechermann_multiorbital_2020,lee_infinite-layer_2004,sakakibara_model_2019,zhang_effective_2019,hepting_electronic_2020,wu_robust_2020,karp_many-body_2020,been_theory_2020,gu_substantial_2020}, DFT+DMFT \cite{wang_hunds_2020,kitatani_nickelate_2020}, and one-shot G$_0$W$_0$ \cite{olevano_ab_2020}, the total spectral function shows that there are two bands crossing the Fermi level. Of these two bands, the lower energy band shows strong two dimensional character, and it is dominated by the Ni-$d_{x^2-y^2}$ orbital. The remaining band, the so called self-doping band, is the higher energy  band which shows strong hybridization between other Ni orbitals and La orbitals. The band dispersion of the self-doping band varies strongly along the direction normal to the Ni-O plane ($\hat{z}$), demonstrating its strong 3-dimensional character \cite{hepting_electronic_2020}. Moreover, the orbital character of the self-doping band is strongly dependent on $k_z$. In the $k_z$=0 plane, the orbital character of the self-doping band is mostly La-$d_{z^2}$ and Ni-$d_{z^2}$ \cite{lee_infinite-layer_2004,lechermann_late_2020}, In contrast, in the $k_z=\pi/c$ plane, where $c$ is the lattice constant along the $\hat{z}$ direction, its orbital character is mostly La-$d_{xy}$ and Ni-$p_{z}$. This analysis is consistent with a recent two band model study \textit{from first-principles}, showing that the two Fermi-level-crossing bands can be spanned by a Ni-$d_{x^2-y^2}$ orbital and an axial orbital \cite{adhikary_orbital_2020}. The axial orbital is not centered on a single atom. Instead, its density is centered on both the Ni and La atoms.

\begin{table}[t]%
  \caption{Electron occupation of Ni-$d$ orbitals in La$_{0.8}$Ba$_{0.2}$NiO$_2$ and Fe-$d$ orbitals in FeSe at T=300K}
  \begin{tabularx}{0.95\columnwidth}{c c *{6}{Y}}
    \hline
    \hline    
    & Materials                   & $d_{xy}$ & $d_{yz}$ & $d_{xz}$ & $d_{z^2}$ & $d_{x^2-y^2}$\\
    \hline

    & La$_{0.8}$Ba$_{0.2}$NiO$_2$ & 1.94     & 1.89     & 1.89     & 1.59     & 1.04          \\ 
    & FeSe                        & 1.22     & 1.19     & 1.19     & 1.44      & 1.26          \\


    
    \hline
    \hline    
  \end{tabularx}
  \label{tab_orb}
\end{table}

Orbital occupation in the Ni-$d$ orbitals differentiates the $t_{2g}$ and $e_g$ orbitals. 
As summarized in Table \ref{tab_orb}, the Ni-$e_{g}$ orbitals are partially filled but the Ni-$t_{2g}$ orbitals are fully-filled in La$_{0.8}$Ba$_{0.2}$NiO$_2$. This orbital occupation profile is far from a prediction based on oxidation state rules, \textit{i.e.}, 2, 2, 2, 2, and 1 for Ni-$d_{xy}$, Ni-$d_{yz}$, Ni-$d_{xz}$, Ni-$d_{z^2}$, and Ni-$d_{x^2-y^2}$, respectively. Intriguingly, the difference stands out especially for the  Ni-$z^2$ orbital, which is far from the expected double occupation \cite{leonov_lifshitz_2020,petocchi_normal_2020}. This discrepancy can be explained by the hybridization between Ni-d$_{z^2}$ and La-$d_{z^2}$ orbitals. The strong hybridization between these two orbitals in the $\Gamma$-$X$-$M$-$\Gamma$ plane makes the Ni-d$_{z^2}$ orbital exhibit a dispersion which is distinct from its dispersion in isolation (the flat band at E$_F$-1eV in the $Z$-$R$-$A$-$Z$ plane in Fig. \ref{fig_spectral}(c)) \cite{choi_quantum-fluctuation-frustrated_2020}. Indeed, upon Ba doping up to 0.3, only $\sim$ 25\% of the added holes go into the Ni-$d$ orbitals, while  the remaining holes go into other orbitals, especially the La-$d_{xy}$, La-$d_z$ and Ni-$p_z$ orbitals (as shown in the supplementary materials). This is consistent with other theoretical studies at low-doping \cite{leonov_lifshitz_2020,wang_hunds_2020}, and it makes the $t_{2g}$-$e_g$ differentiation in orbital occupation robust against low extrinsic hole-doping. Here we note that the orbital occupation as well as the orbital resolved spectral functions are dependent on the choice of the Wannier orbitals. To construct atomic-orbital-like Wannier orbitals tightly bounded and centered on the atoms, we constructed 31 atom-centered Wannier orbitals for each spin (see the supplementary materials).
\begin{figure}[t]
  \centering
  \includegraphics[width=1.00\columnwidth]{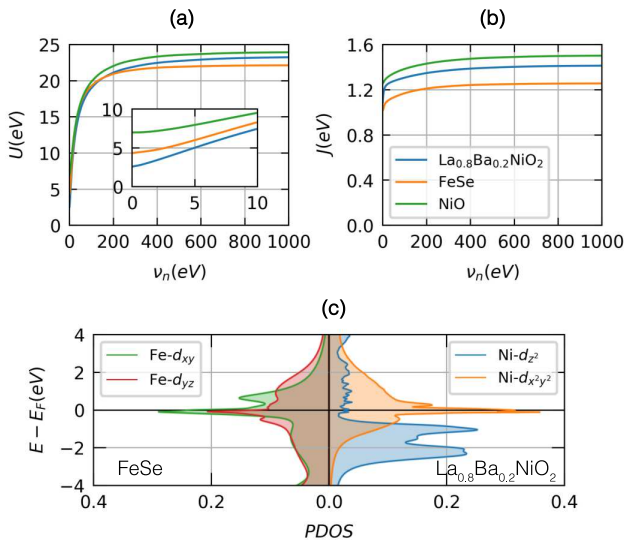}
  \caption{(a) $U$ and (b) $J$ for Ni-$d$ orbitals in La$_{0.8}$Ba$_{0.2}$NiO$_2$, Fe-$d$ orbitals in FeSe, and Ni-$d$ orbitals in NiO within the constrained random phase approximation. In the static limit, the $U$ of the Ni-$d$ orbitals in La$_{0.8}$Ba$_{0.2}$NiO$_2$ is much smaller than in NiO and even smaller than that of the Fe-$d$ orbitals in FeSe. In the entire frequency range, the $J$ for Ni-$d$ orbitals in La$_{0.8}$Ba$_{0.2}$NiO$_2$ is larger than the $J$ of Fe-$d$ orbitals in FeSe. (c) Projected density of states to La-$e_g$ orbitals in La$_{0.8}$Ba$_{0.2}$NiO$_2$ and Fe-$t_{2g}$ orbitals in FeSe}
  \label{fig_pdos}
\end{figure}

Based on the Coulomb interaction calculation within the constrained random phase approximation (cRPA), it is legitimate to assume the dominance of Hundness over ``Mottness'' in La$_{1-\delta}$Ba$_{\delta}$NiO$_2$. Fig. \ref{fig_pdos} shows the on-site Hubbard and Hund interactions among five Ni-$d$ orbitals within the constrained random phase approximation. For comparison, we plotted the $U$ and $J$ of Ni-$d$ orbitals in NiO and Fe-$d$ orbitals in FeSe. As is typical, $U$ is strongly frequency-dependent, while $J$ is not. Interestingly, the static $U$ of the Ni-$d$ orbitals in La$_{0.8}$Ba$_{0.2}$NiO$_2$ is much smaller than it is in the charge-transfer insulator NiO. It is even smaller than the $U$ of Fe-$d$ orbitals in the Hund's metal FeSe. In contrast, the $J$ of the Ni-$d$ orbitals in La$_{0.8}$Ba$_{0.2}$NiO$_2$ is even larger than the $J$ of Fe-$d$ in the Hund's metal FeSe. Judging from the fact that the Ni-$e_g$ orbitals in La$_{1-\delta}$Ba$_{\delta}$NiO$_2$ and the Fe-$t_{2g}$ orbitals in FeSe show similar bandwidths, we can safely assume the dominant role of Hundness over Mottness in La$_{1-\delta}$Ba$_{\delta}$NiO$_2$.

\begin{figure}[t]
  \centering
  \includegraphics[width=0.95\columnwidth]{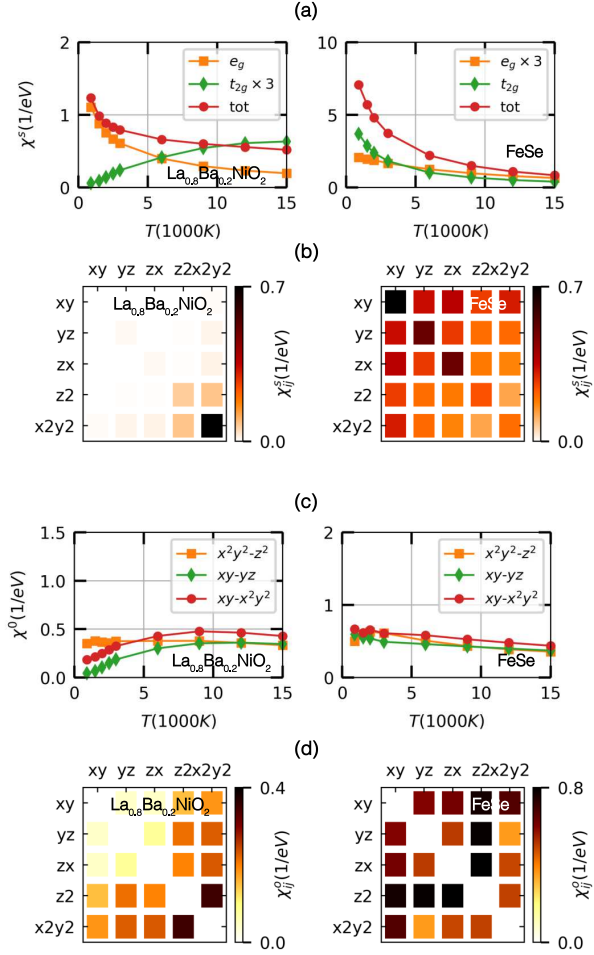}
  \caption{The temperature dependence of the local spectrum of the spin and orbital susceptibilities. (a) The temperature dependence of static spin susceptibility ($\chi^s$) of $d$ orbitals (red dots), $t_{2g}$ orbitals (green diamonds), and $e_g$-orbitals (orange squares) in La$_{0.8}$Ba$_{0.2}$NiO$_2$ and FeSe. (b) Orbital-resolved static spin susceptibility ($\chi_{ij}^s$) of Ni-$d$ orbitals in La$_{0.8}$Ba$_{0.2}$NiO$_2$ and Fe-$d$ orbitals in FeSe at T = 900 K. (c) The temperature dependence of static orbital susceptibility ($\chi_{ij}^o$) of Ni-$d$ orbitals in La$_{0.8}$Ba$_{0.2}$NiO$_2$ and Fe-$d$ orbitals in FeSe. (d) Orbital susceptibility ($\chi_{ij}^s$) of Ni-$d$ orbitals in La$_{0.8}$Ba$_{0.2}$NiO$_2$ and Fe-$d$ orbitals in FeSe at T = 900 K.}
  \label{fig_sus}
\end{figure}

To understand the origin of strong correlations in the infinite-layer nickelates further, we calculated the temperature and doping dependence of Ni-$d_{x^2-y^2}$ local spectra as well as static spin- and orbital-susceptibility. These one- and two-particle quantities are ``litmus-papers'' to quantify the relative strength of Hundness versus Mottness. Hund's metals show various characteristic behaviors. One is spin-orbital separation: a two-step screening process in which local spin moment is screened at much lower temperature than local orbital polarization. Another is the absence of the pseudo gap in the local spectra. At high temperature when quasiparticle spectral weight near the Fermi level is transferred into high-energy Hubbard bands, spectral weight at the Fermi level is still not negligible and the local spectra is dominated by a single incoherent peak. In contrast, in the correlated metallic system where Mottness dominates, spin-orbit separation is negligible. In addition, the high-temperature spectral weight at the Fermi level is depleted due to the quasiparticle spectral weight transfer and pseudogap forms at the Fermi level at the high temperature. By calculating these quantities, we found multiple Hundness signatures. More importantly, these signatures are primarily evident in the active Ni-$e_g$ orbitals and not the inactive Ni-$t_{2g}$ orbitals. 

Five Ni-$d$ orbitals in La$_{1\mhyphen\delta}$Ba$_{\delta}$NiO$_2$ show clear spin-orbital separation. Fig. \ref{fig_sus}(a) and Fig. \ref{fig_sus}(c) show the temperature dependence of the static local susceptibility in spin ($\chi_{tot}^s$) and orbital ($\chi_{ij}^o$) channels. These are defined as $\chi_{tot}^s=\sum_{ij=d}\chi_{ij}^s, \chi_{ij}^s=\int_0^\beta d\tau \langle S_{iz}(\tau)S_{jz}(0)\rangle$, and $\chi_{ij}^o=\int_0^\beta d\tau \langle N_{i}(\tau) N_{j}(0)\rangle-\beta\langle N_{i} \rangle \langle N_{j}\rangle$. Here $S_{iz}(\tau)$ is the orbital-resolved spin operator and $N_i$ is the orbital resolved occupation operator. According to Deng \textit{et al.} \cite{deng_signatures_2019}, the temperatures at which the screening  of the spin and orbital degrees of freedom becomes noticeable are one of the key measures with which to distinguish between Mott and Hund physics. These onset screening temperatures in spin and orbital channels can be obtained by estimating the temperature at which these susceptibilities deviates from Curie-like behaviors. In the Mott regime, these two energy scales coincide. In contrast, in the Hund regime, the orbital onset temperature is much higher than the spin onset temperature. At a temperature between these two onset temperatures, the spin susceptibility is Curie-like but the orbital-susceptibility is Pauli-like. This is exactly the behavior seen in FeSe. In FeSe, the local spin susceptibility is Curie-like (red dots in Fig. \ref{fig_sus}(a)), but the local orbital susceptibility approaches its maximum upon cooling (red dots in Fig. \ref{fig_sus}(c)). La$_{0.8}$Ba$_{0.2}$NiO$_2$ behaves like FeSe. The spin degree of freedom (red dots in Fig. \ref{fig_sus}(a)) shows Curie-like behavior. In contrast, the orbital susceptibility between any Ni-$d$ orbital pair shows a downturn of the susceptibility upon cooling (red dots in Fig. \ref{fig_sus}(c)). 


\begin{figure*}[t]
  \centering
  \includegraphics[width=0.9\textwidth]{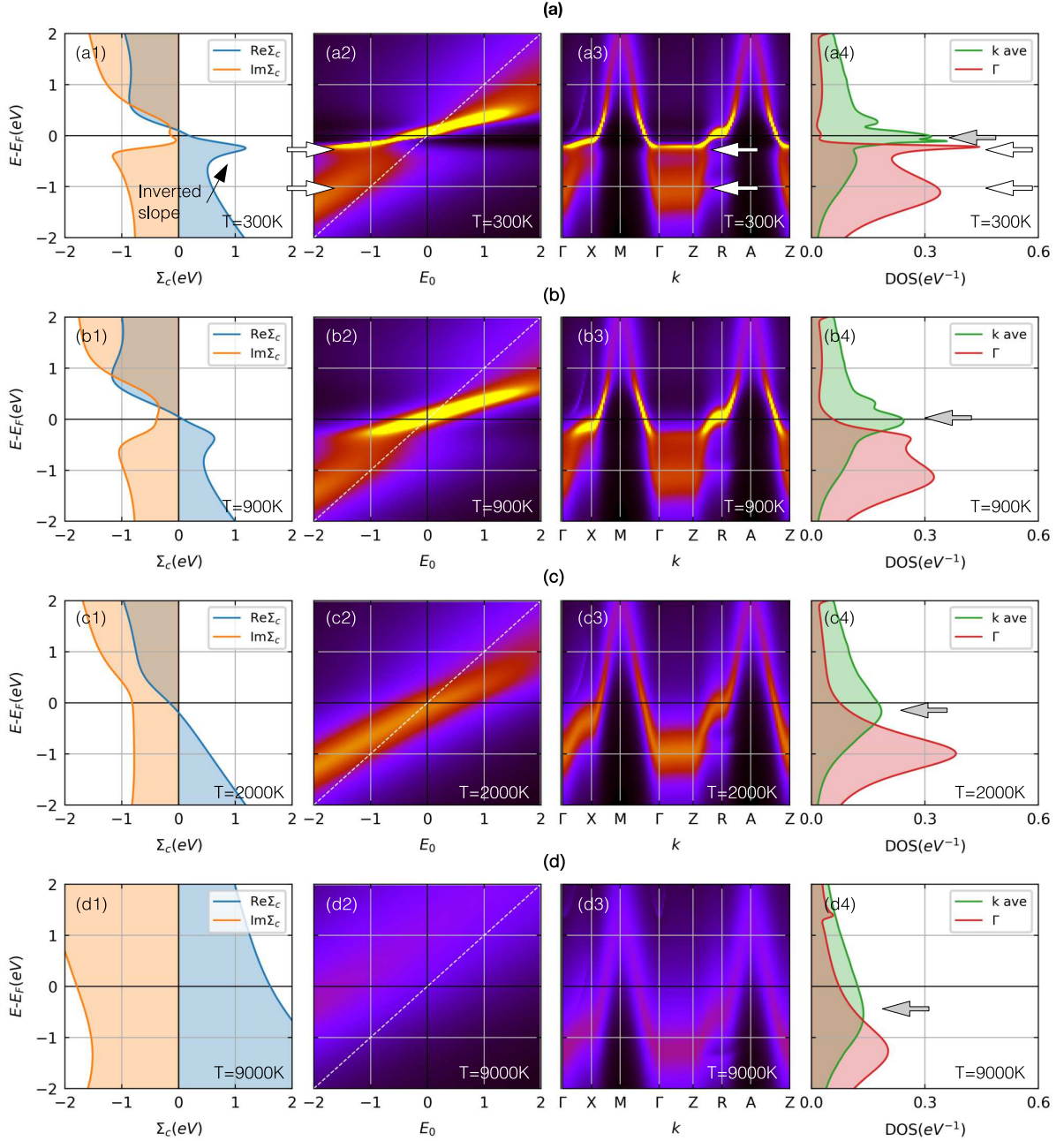}
  \caption{Spectral data obtained for Ni-$d_{x^2-y^2}$ orbital in La$_{0.8}$Ba$_{0.2}$NiO$_2$ at four different temperatures of T=300K (the first row), T=900K (the second row), T=2000K (the third row), and T=9000K (the third row). First column: electron correlation self-energy ($\Sigma_c$). Second column: spectral function of an auxiliary Green's function of $A(E_0, E)=-\frac{1}{\pi}Im\left(\frac{1}{E-E_0-\Sigma_c(E)}\right)$. White dashed line shows the dispersion of the bare band of $E=E_0$. Third column: orbital resolved spectral function. Fourth column: orbital-resolved density of states and orbital-resolved spectral function at $\Gamma$ point. White arrows in the second, third and fourth columns indicate the energies of the two peaks in the orbital-resolved spectral function at the $\Gamma$ point. Gray arrow in the fourth column indicates the peak in the orbital-resolved density of states. } 
  \label{fig_waterfall}
\end{figure*}  


However, there is an important distinction between the Ni-$d$ orbitals in La$_{1\mhyphen\delta}$Ba$_{\delta}$NiO$_2$ and Fe-$d$ orbitals in FeSe: The $t_{2g}$ orbitals in La$_{1\mhyphen\delta}$Ba$_{\delta}$NiO$_2$ are inactive. In spite that Ni-$t_{2g}$ is almost fully filled in La$_{1\mhyphen\delta}$Ba$_{\delta}$NiO$_2$, the inactivity of Ni-$t_{2g}$ orbitals for the Hundness-related two-particle quantities ($\chi_{ij}^s$ and $\chi_{ij}^o$) is a non-trivial question. Inactivity in the one-particle level (single particle Green's function) is not sufficient to assure inactivity in the two-particle level (the local susceptibilities). This can be illustrated by the charge susceptibility data obtained within multitier GW+EDMFT by F. Petocchi et al. \cite{petocchi_normal_2020}. As shown in Fig. 3 of the paper, the intraorbital charge fluctuation associated with Ni-$d_{xz/yz}$ orbitals is not negligible but comparable to the fluctuation associated with Ni-$d_{x^2-y^2}$  although Ni-d$_{xz/yz}$ orbital is almost fully-filled within their approach. To convince the inactivity of Ni-$t_{2g}$ orbitals in the two particle level, their Hundness-related two-particle quantities ($\chi_{ij}^s$ and $\chi_{ij}^o$) should be examined explicitly. 

First, spin fluctuations are not active among the Ni-$t_{2g}$ orbitals. Fig. \ref{fig_sus} (b) shows $\chi_{ij}^s$. In FeSe, all possible pairs of Fe-$d$ orbitals show a strong spin response. In contrast, only the Ni-$e_g$ subspace exhibits a strong spin response in La$_{0.8}$Ba$_{0.2}$NiO$_2$, while the response due to the remaining pairs is strongly suppressed. The temperature dependence of the spin susceptibility in the t$_{2g}$ subspace ($\chi_{t_{2g}}^s$) further supports the distinction between the Ni-$d$ orbital and Fe-$d$ orbitals. Here, $\chi_{t_{2g}}^s=\sum_{ij=t_{2g}}\chi_{ij}^s$. As shown in Fig. \ref{fig_sus} (a), $\chi_{t_{2g}}^s$ (green diamonds) in La$_{0.8}$Ba$_{0.2}$NiO$_2$ strongly deviates from the Curie-like behaviors of $\chi_{tot}^s$. This does not occur in FeSe. Most strikingly, $\chi_{t_{2g}}^s$ approaches zero upon cooling.



Second, the static orbital susceptibility shows the suppression of orbital fluctuations in the Ni-t$_{2g}$ subspace. Fig. \ref{fig_sus}(d) shows $\chi_{ij}^o$. In FeSe, all possible pairs of Fe-$d$ orbitals show a strong orbital response. In contrast, the $\chi_{ij}^o$ in the Ni-$t_{2g}$ subspace are strongly suppressed in La$_{0.8}$Ba$_{0.2}$NiO$_2$. The temperature dependence of the orbital susceptibility  in the t$_{2g}$ subspace ($\chi_{xy,yz}^o$), shown in Fig. \ref{fig_sus} (c), is another distinction between Ni-$d$ orbitals and F-$d$ orbitals. Here, in contrast to FeSe, where $\chi_{xy,yz}^o$ (green diamonds) follows $\chi_{x^2\text{-}y^2,z^2}^o$ (orange squares), $\chi_{xy,yz}^o$ (green diamonds) in La$_{0.8}$Ba$_{0.2}$NiO$_2$ strongly deviates from $\chi_{x^2\text{-}y^2,z^2}^o$(orange squares). Most strikingly, $\chi_{xy,yz}^o$ approaches zero upon cooling.


Once we narrow down our view from all Ni-$d$ orbitals into only the Ni-$e_g$ orbitals, we can successfully find all signatures of a Hund's metal. Two Ni-$e_g$ orbitals in La$_{1\mhyphen\delta}$Ba$_{\delta}$NiO$_2$ show clear spin-orbital separation. Fig. \ref{fig_sus}(a) and Fig. \ref{fig_sus}(c) depict the temperature dependence of static local spin ($\chi_{e_g}^s$) and orbital ($\chi_{x^2-y^2,z^2}^o$) susceptibility.  Here $\chi_{e_g}^s=\sum_{ij=e_{g}} \chi_{ij}^s$. $\chi_{e_g}^s$ (orange squares in Fig. \ref{fig_sus}(a)) shows Curie-like temperature dependence but $\chi_{x^2-y^2,z^2}^o$ (orange squares in Fig. \ref{fig_sus}(c)) shows Pauli-like temperature dependence.

Here we note that there are two more characteristic phenomena of Hund's metal. One is the spin freezing phase \cite{werner_spin_2008}. At a temperature where orbital fluctuation is screened but spin flucuation is not, spin fluctuation doesn't decay to zero at long imaginary time ($\tau=\beta/2$), where $\beta$ is inverse temperature. The other is orbital-decoupling: the suppression of the instantaneous interorbital charge fluctuation\cite{de_medici_hunds_2011}. In these two quantities, we also found evidances of Ni-$e_g$ Hundness. Please see the supplementary materials. 


Hund's physics in the infinite-layer nickelates can be tested further by measuring the temperature evolution of the Ni-$d_{x^2-y^2}$-orbital-resolved spectral function, which dominates the spectra at the Fermi level. According to Deng et al.\cite{deng_signatures_2019}, the high-temperature spectra of the orbital-resolved density of states can be used to confirm Hund's metal physics. At low temperature, spectral weight at the Fermi level is dominated by quasiparticle resonance peak in both Hund's and Mott's metallic phase. However, at a high temperature when quasiparticle spectral weight at the Fermi level are transferred to high-energy Hubbard bands, local spectra distinguish Hund-like and Mott-like metallic systems. In the metallic phase where Mott features dominate, the upper and lower Hubbard bands are well separated from each other due to its proximity to Hubbard-Mott transition and pseudo-gap forms. In contrast, in Hund's metallic phase, the upper Hubbard band is overlapping with the lower Hubbard band and the whole spectra is dominated by a single incoherent peak that has a large value at the Fermi level. This Hund's metallic features are accompanied by shoulder-like structure in the electron self-energy imaginary part as well as the inverted slope of the self-energy real part near the Fermi level \cite{stadler_dynamical_2015,stadler_hundness_2019}. 

Fig. \ref{fig_waterfall} shows the temperature evolution of the Ni-$d_{x^2-y^2}$-orbital-resolved spectral function of La$_{0.8}$Ba$_{0.2}$NiO$_2$. Here we note that the estimated onset screening temperature  in the spin and orbital channels are 3000K and 35000K, respectively. Importantly, up to T=15000K, we were not able to observe pseudo gap formation in the Ni-$d_{x^2-y^2}$ projected density of states. Instead, the local spectrum is composed of a single incoherent peak that has a large value at the Fermi level. In addition, the center of the incoherent peak moves away from the Fermi-level upon heating. Furthermore, the correlation part of the electronic self-energy shows expected Hund's metallic behaviors. As shown in Fig. \ref{fig_waterfall} (a1) there is a shoulder-like structure in its imaginary part self-energy at T=300K. The slope of the real part self-energy is inverted accordingly.  To check its role in the spectral properties, we constructed an auxiliary Green's function of $G_{aux}(E_0, E)=\frac{1}{E-E_0-\Sigma_c(E)}$, which is often used to study Hund's metal physics in the various Hund-Hubbard models \cite{stadler_hundness_2019,kugler_orbital_2019}. Due to the shoulder-like structure in the electron self-energy, the band structure of the auxiliary system is strongly renormalized with a renormalization factor of ~0.2 near the Fermi level. Furthermore, at the negative bare energy ($E_0$), there is strong redistribution of the spectral weight, resulting in an additional incoherent peak. This creates the waterfall features in the Ni-$d_{x^2-y^2}$ orbital resolved spectral function in real materials. As shown in Fig. \ref{fig_waterfall} (a3), the spectral weight along the $\Gamma$-$Z$ line is split into strongly renormalized coherent peak and incoherent peak. As the temperature increases, the shoulder-like structure in the imaginary part of the self-energy becomes weaker. Subsequently, the coherenet and the coherent peaks merge.

\begin{figure}[t]
  \centering
  \includegraphics[width=1.0\columnwidth]{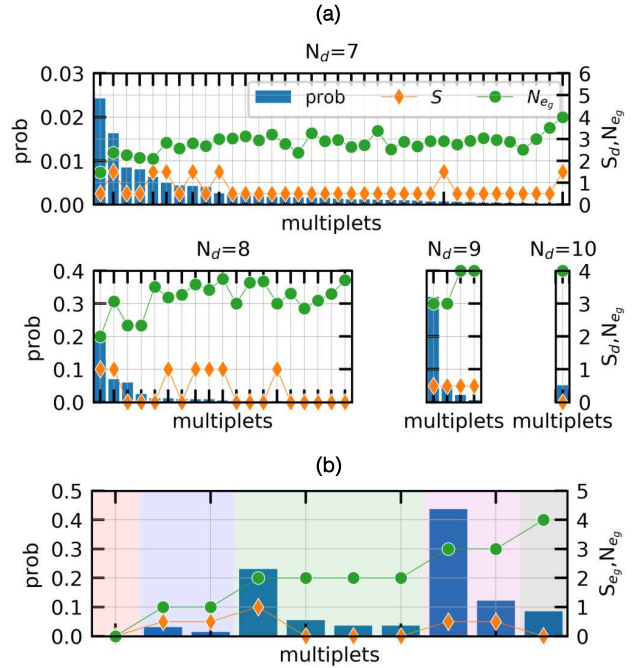}
  \caption{ (a) Reduced local many-body density on the Ni-3$d$ multiplets in La$_{0.8}$Ba$_{0.2}$NiO$_2$ at T=300K. Each multiplet has been labeled by using the Ni-3$d$ total spin ($S_d$) and Ni-3$d$ total charge ($N_d$). The Ni-$e_g$ total charge ($N_{e_g}$) are also shown. (e) Reduced local many-body density on Ni-$e_g$ multiplets in La$_{0.8}$Ba$_{0.2}$NiO$_2$ at T=300K. Each multiplet has been labeled by using Ni-$e_g$ total spin ($S_{e_g}$), Ni-$e_g$ total charge ($N_{e_g}$) and atomic-limit eigenenergy.}
  \label{fig_histo}
\end{figure}

To clarify the microscopic origin of Ni-$e_g$ Hund's metallic behaviors, we investigate the reduced local many-body density or local probabilities of Ni-3$d$ multiplet states in the atomic limit. Fig. \ref{fig_histo} (a) shows the valence histogram for the Ni-3$d$ multiplets in La$_{0.8}$Ba$_{0.2}$NiO$_2$. That is, it shows a partial trace of the density matrix of the full Hilbert space, where this partial trace excludes the Ni-3$d$ subsystem in order to reveal the probability that a given multiplet state in the correlated Ni-3$d$ subsystem is occupied. It is traced further over the secondary spin quantum number. We decompose the Ni-3$d$ subspace according to the total charge ($N_d$) of the mutliplet states, and find that for $N_d=$7, 8, 9 and 10, the most probable states involve the total spin $S_d=1/2$, $1$, $1/2$, and $0$ as well as the occupation of the  $e_g$ orbitals ($N_{e_g}$) is 1.47. 2, 3, and 4, respectively. Interestingly, these can be interpreted as the multiplets which maximize the total spin of the Ni-$e_g$ electron in each $N_{e_g}$ subspace; these  are not the multiplets which maximize the total spin of all N-3$d$ electrons in each $N_d$ subspace. The reduced local many-body density on the Ni-$e_g$ multiplets shown in Fig. \ref{fig_histo} (b) supports this observation. The most probable Ni-$e_g$ multiplet in each $N_{e_g}$ subspace is the one with maximum Ni-$e_g$ total spin ($S_{e_g}$).  Again, this supports our conclusion that Hund metallic behaviors are limited to the Ni-$e_g$ orbitals.

In addition to onsite Hund's coupling, the crystal-field splitting between Ni-$d_{z^2}$ and Ni-$d_{x^2-y^2}$ orbitals is another important factor to control the physical quantities to judge Hundness versus Mottness. The crystal field splitting plays a two-faced role in those quantities. On one hand, it amplifies Hundness signatures. To illustrate, the non-zero crystal field splitting suppresses spin Kondo temperature but enhances orbital Kondo temperatures, thus boosting spin-orbital separation \cite{kugler_orbital_2019}. Thus, the spin-orbital separation in the system with a non-zero crystal-field can not be the signature of Hundness. On the other hand, it enhances Mottness signatures. The crystal field splitting makes possible Ni-$e_g$ spin-singlet lower in energy than the spin-triplet states \cite{werner_high-spin_2007}. It also increases the separation between lower and upper Hubbard bands, thus promoting pseudo-gap formation. The enhancement of the Mottness signatures can be understood by using the Kanamori Hamiltonian in its atomic limit. By assuming inactivity Ni-$t_{2g}$ orbitals, the local physics at the Ni site may be understood by the $e_g$-Kanamori Hamiltonian. In its atomic limit with vanishing intersite hopping, the Hamiltonian is given by

\begin{equation}
  \begin{split}
    H &=  -\Delta \sum_\sigma n_{2\sigma}+U\sum_{i}{n_{i \uparrow} n_{i \downarrow}} \\
    &+ \sum_{i,j,\sigma,\sigma'}^{i \neq j}\left(U'-J\delta_{\sigma,\sigma'}\right) n_{ i \sigma} n_{ j \sigma'} \\
    &- J\sum_{i,j}^{i \neq j}\left(c_{i \uparrow}^{\dagger} c_{i \downarrow} c^{\dagger}_{j \downarrow} c_{j \uparrow} 
      - c^{\dagger}_{ i \uparrow} c^{\dagger}_{ i \downarrow} c_{ j \downarrow} c_{ j \uparrow}\right) , 
    \label{eq_kanamori}
  \end{split}
\end{equation}
Here, $\Delta$, $U$, $U'$ and $J$ are the crystal-field splitting which is positive, intraorbital Coulomb interaction, interorbital Coulomb interaction, and Hund's coupling, respectively. 
When $\Delta=0$, triplet states are always the lowest-energy states in N$_{e_g}$= 2 subspace. However, non-zero crystal-field splitting enables the singlet ground states formation in $N_{e_g}=2$ subspace when $\Delta >\sqrt{(U-U'+J)^2-J^2}$. Here we note that $U>U'$ in the realistic materials. Furthermore, the $\Delta$ promotes pseudo-gap formation by enhancing the separation between upper and lower Hubbard bands in the weakly hole-doped regime from $N_{e_g}=3$ filling. The separation ($U^{eff}$) is given by $U^{eff}=U^{eff}|_{\Delta=0}+(2N_{e_g}-5)\Delta$ when triplet states are the ground states in $N_{e_g}=2$ subspace and $U^{eff}=U^{eff}|_{\Delta=0}+(2N_{e_g}-5)\Delta-(3N_{e_g}-8)(\sqrt{J^2+\Delta^2}-J) $ when a singlet is the ground state in $N_{e_g}=2$ subspace. Here $U^{eff}|_{\Delta=0}$ is the energy gap when $\Delta$=0.  $U^{eff}\geq U^{eff}|_{\Delta=0}$ in the electron occupation of $2.5<N_{e_g}<3$ regardless of $N_{e_g}=2$ subspace ground state. For the derivation, please see the supplementary materials.

Despite the crystal-field-induced enhancement of the pseudo-gap as well as singlet population, both measures advocate Hund's metallicity of La$_{1-\delta}$Ba$_{\delta}$NiO$_2$ as shown in Fig. \ref{fig_histo}(b) and Fig. \ref{fig_waterfall}. Together with the spin-orbital separation shown in Fig. \ref{fig_sus}, these signatures indicate that La$_{1-\delta}$Ba$_{\delta}$NiO$_2$ is a strong candidate of two-orbital Hund's metal.

\begin{figure}[t]
  \centering
  \includegraphics[width=0.95\columnwidth]{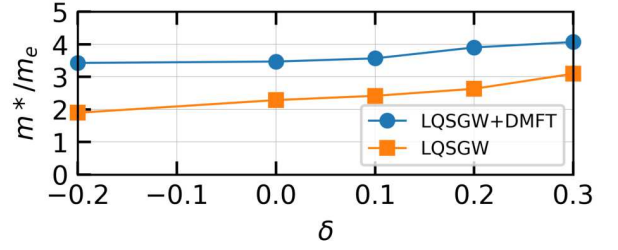}
  \caption{Doping dependence of Ni-$d_{x^2-y^2}$ band cyclotron effective mass in the $k_z=0$ plane within LQSGW+DMFT (blue, T=150K) and LQSGW (orange) methods. $m_e$ denotes the free electron mass.}
  \label{fig_mass}
\end{figure}

Lastly, we propose another experiment to support Ni-$e_g$ Hundness in the infinite-layer nickelates: the doping dependence of Ni-$d_{x^2-y^2}$ band effective mass. In a paramagnetic system where the proximity to Mott transition dominates electron correlation and single-band is a good minimum model to describe the low-energy physics, the effective mass is expected to be maximum in the undoped system and decreases if the system is either hole-doped or electron-doped. In contrast, as demonstrated by the Fe-based superconductors \cite{de_medici_selective_2014}, the effective mass of Hund's metals changes monotonically from hole-doped side to electron-doped side in Hund's metals. Fig. \ref{fig_mass} shows the doping dependence of the cyclotron effective mass of the Ni-$d_{x^2-y^2}$ bands in the $k_z=0$ plane. Both LQSGW+DMFT and LQSGW methods show that the effective mass increases monotonically from electron doped side to hole-doped side. This monotonic doping dependence of the effective mass could be confimed by other experiments such as specific heat measurement as well as angle-resolved photoemission spectroscopy. In contrast to other signatures proposed in this paper, the doping dependence of the Ni-$d_{x^2-y^2}$ band effective mass does not require high temperature measurements.

\textit{Conclusion.} By using \textit{ab initio} LQSGW+DMFT methodology, we demonstrated that on-site Hund's coupling in Ni-$d$ orbitals results in multiple signatures of Hund's metallic phase in Ni-$e_g$ orbitals. Our finding sheds a new light on Hundness in the correlated quantum materials and has potential implications for the broad range of correlated two orbital systems away from half-filling and the role of on-site Hund's coupling \cite{stadler_model_2019,de_medici_hunds_2011,werner_nickelate_2020}. 


\textit{Acknowledgments}. S.C. thanks G. L. Pascut, and C-.J-. Kang and for useful conversation. This work was supported by the U.S Department of Energy, Office of Science, Basic Energy Sciences as a part of the Computational Materials Science Program. S. R and M. J. H were supported by NRF Korea (2018R1A2B2005204 and 2018M3D1A1058754). This research used resources of the National Energy Research Scientific Computing Center (NERSC), a U.S. Department of Energy Office of Science User Facility operated under Contract No. DE-AC02-05CH11231.

\bibliography{zotero}

\end{document}


\title{Supplementary materials for \\
  ``Infinite-layer nickelates as Ni-$e_g$ Hund's metals''}
\date{\today}

\author{Byungkyun Kang}
\affiliation{
  Condensed Matter Physics and Materials Science Department, 
  Brookhaven National Laboratory, Upton, NY 11973, USA
}
\author{Corey Melnick}
\affiliation{
  Condensed Matter Physics and Materials Science Department, 
  Brookhaven National Laboratory, Upton, NY 11973, USA
}
\author{Patrick Semon}
\affiliation{
  Condensed Matter Physics and Materials Science Department, 
  Brookhaven National Laboratory, Upton, NY 11973, USA
}
 \author{Siheon Ryee}
 \affiliation{Department of Physics, KAIST, Daejeon 34141, Republic of Korea}
 \author{Myung Joon Han}
 \affiliation{Department of Physics, KAIST, Daejeon 34141, Republic of Korea}

\author{Gabriel Kotliar}
\affiliation{
  Department of Physics and Astronomy, Rutgers University, New Jersey 08854, USA
}
\affiliation{
  Condensed Matter Physics and Materials Science Department, 
  Brookhaven National Laboratory, Upton, NY 11973, USA
}

\author{Sangkook Choi}
\email[E-mail me at:]{sachoi@bnl.gov}
\affiliation{
  Condensed Matter Physics and Materials Science Department, 
  Brookhaven National Laboratory, Upton, NY 11973, USA
}

\maketitle
\section{LQSGW calculation}
LQSGW calculations are performed by using FlapwMBPT package \cite{kutepov_linearized_2017}, which is based on full-potential linearized augmented plane wave plus local orbital method. For La$_{1-\delta}$Ba$_{\delta}$NiO$_2$, experimental lattice constants and atomic positions \cite{hayward_synthesis_2003} for NdNiO$_2$ are used. The Muffin-tin (MT) radius ($R$) is selected as follows: 2.7 for La/Ba, 2.1 for Ni, and 1.8 for O in Bohr radius. Wave functions are spanned by spherical harmonics with $l$ up to 8 for La/Ba, 6 for Cu, and 6 for O in the MT spheres. In the interstitial region (IS), it is spanned by plane waves with the cutoff ($K_{cut}$) of $R_{Ni}$×$K_{cut}$= 8.8. Product basis set is spanned by spherical harmonics with l up to 8 for La/Ba, 6 for Cu, and 6 for O in the MT spheres and by planewaves with the cutoff ($G_{cut}$) of $R_{Ni}$×$G_{cut}$= 13 in IS region. All the unoccupied states are taken into account for both polarizability and self-energy calculation. The Brillouin zone is sampled in 4×4×4 grid.

For FeSe and NiO, please see this paper \cite{choi_comdmft:_2019}.



\section{Wannier function constructions}
For La$_{1-\delta}$Ba$_{\delta}$NiO$_2$, 31 Wannier functions are constructed with a frozen energy window between -10 eV to 11.2 eV and with a disentanglement energy window of -10 eV to 52 eV: La-s, La-p, La-d, La-f, Ni-s, Ni-p, Ni-d, and O-p orbitals. Initial trial orbitals are constructed by using Muffin-tin orbitals in LAPW basis set with well-defined angular momentum characters. Final Wannier orbitals are centered exactly at the targeted atom. Their orbital characters, calculated Wannier centers, and spreads are in Table \ref{tab_wan}. 
\begin{table}[H]%
  \centering
  \caption{Wannier function orbital characters, calculated wannier centers, and spreads of La$_{0.8}$Ba$_{0.2}$NiO$_2$} 
  \begin{tabularx}{0.8\textwidth}{l c *{1}{Y} c *{1}{Y} c *{1}{Y}}
    \hline
    \hline 
    & orbital & type & center & spread ($\AA^2$) \\
    \hline  

    & 1 &La-$s$  &( 1.960000, 1.960000, 1.640000 ) & 2.74772016\\                         
    & 2 &La-$p_y$    &( 1.960000, 1.960000, 1.640000 )&  2.62769922\\
    & 3 &La-$p_z$    & ( 1.960000, 1.960000, 1.640000 )&  2.44701728\\
    & 4 &La-$p_x$    &( 1.960000, 1.960000, 1.640000 )&  2.62769922\\
    & 5 &La-$d_{xy}$    &( 1.960000, 1.960000, 1.640000 )&  2.14236838\\
    & 6 &La-$d_{yz}$    &( 1.960000, 1.960000, 1.640000 )&  1.89698543\\
    & 7 &La-$d_{z^2}$    &( 1.960000, 1.960000, 1.640000 )&  2.01022884\\
    & 8 &La-$d_{xz}$    &( 1.960000, 1.960000, 1.640000 )&  1.89698543\\
    & 9 &La-$d_{x^2-y^2}$   &( 1.960000, 1.960000, 1.640000 )&  1.95727422\\
    &10 &La-$f_{y(3x^2-y^2)}$ &   ( 1.960000, 1.960000, 1.640000 )&  1.08079931\\
    &11 &La-$f_{xyz}$   & ( 1.960000, 1.960000, 1.640000 )&  1.16139012\\
    &12 &La-$f_{yz^2}$   &( 1.960000, 1.960000, 1.640000 )&  1.00988505\\
    &13 &La-$f_{z^3}$   &( 1.960000, 1.960000, 1.640000 )&  0.84709070\\
    &14 &La-$f_{xz^2}$   &( 1.960000, 1.960000, 1.640000 )&  1.00988505\\
    &15 &La-$f_{z(x^2-y^2)}$  &( 1.960000, 1.960000, 1.640000 )&  0.73111829\\
    &16 &La-$f_{x(x^2-3y^2)}$  &( 1.960000, 1.960000, 1.640000 )&  1.08079931\\
    &17 &Ni-$s$  &( 0.000000, -0.000000, 0.000000 )&  1.84519013\\
    &18 &Ni-$p_y$ &( -0.000000, 0.000000, -0.000000 )&  1.57878888\\
    &19 &Ni-$p_z$ &( -0.000000, 0.000000, -0.000000 )&  1.92206829\\
    &20 &Ni-$p_x$ &( 0.000000, -0.000000, -0.000000 )&  1.57878888\\
    &21 &Ni-$d_{xy}$  &( 0.000000, -0.000000, -0.000000 )&  0.39838710\\
    &22 &Ni-$d_{yz}$  &( -0.000000, -0.000000, -0.000000 )&  0.38935269\\
    &23 &Ni-$d_{z^2}$ &( -0.000000, 0.000000, -0.000000 )&  0.34985191\\
    &24 &Ni-$d_{xz}$  &( -0.000000, -0.000000, -0.000000 )&  0.38935269\\
    &25 &Ni-$d_{x^2-y^2}$&( -0.000000, 0.000000, 0.000000 )&  0.31000163\\
    &26 &O-$p_y$&( -0.000000, 1.960000, -0.000000 )&  0.56811085\\
    &27 &O-$p_z$&( -0.000000, 1.960000, -0.000000 )&  0.78206995\\
    &28 &O-$p_x$&( -0.000000, 1.960000, 0.000000 )&  0.82293584\\
    &29 &O-$p_y$&( 1.960000, -0.000000, -0.000000 )&  0.82293584\\
    &30 &O-$p_z$&( 1.960000, -0.000000, -0.000000 )&  0.78206995\\
    &31 &O-$p_x$&( 1.960000, -0.000000, -0.000000 )&  0.56811085\\
    \hline
    \hline 
  \end{tabularx}
  \label{tab_wan}
\end{table}
Here, atomic positions of La, Ni and two O are (1.960000, 1.960000, 1.640000), (0.000000, 0.000000, 0.000000), (0.000000, 1.960000, 0.000000) and (1.960000, 0.000000, 0.000000), respectively. 

\section{Constrained random phase approximation}

We calculated the bosonic Weiss field $\widetilde{\mathcal{U}}$ associated with the correlated $d$ orbitals within constrained random phase approximation (cRPA) \cite{aryasetiawan_frequency-dependent_2004,aryasetiawan_calculations_2006} and its Slater's integrals. Here we stress that the bosonic Weiss field $\widetilde{\mathcal{U}}$ from cRPA is a way to evaluate $\widetilde{\mathcal{U}}$ and not identical to $\widetilde{\mathcal{U}}$ from “ideal” fully self-consistent GW+EDMFT. Within cRPA, the bosonic Weiss field $\widetilde{\mathcal{U}}$ is obtained by separating out the RPA polarizability ($P_{QP}^{low}$) from the correlated states. In ComDMFT $P_{QP}^{low}$ is defined by identifying correlated bands \cite{aryasetiawan_calculations_2006,miyake_screened_2008,werner_satellites_2012}. These are bands having strong Ni-$d$ orbital characters are chosen as correlated bands at each $\mathbf{k}$ point. The number of correlated bands is set to be the number of correlated orbitals for each spin. 

Then, $P_{QP}^{low}$ is defined in the following way. 

\begin{equation}
  \begin{split}
    P_{QP}^{low}(\mathbf{r},\mathbf{r}',\mathbf{k},i\nu_n)=-2\sum_{\mathbf{k'}}\sum_{N}^{\substack{\text{unoccupied }\\\text{correlated bands}}}\sum_{M}^{\substack{\text{occupied}\\\text{correlated bands}}}\\
    \psi_{N\mathbf{k}'}(\mathbf{r})\psi_{M\mathbf{k}'+\mathbf{k}}^*(\mathbf{r})\psi_{N\mathbf{k}'}^*(\mathbf{r}')\psi_{M\mathbf{k}'+\mathbf{k}}(\mathbf{r}')\frac{2(E_{N\mathbf{k}'}-E_{M\mathbf{k}'+\mathbf{k}})}{\nu_n^2-(E_{N\mathbf{k}'}-E_{M\mathbf{k}'+\mathbf{k}})^2},
    \label{eq:chi_low}
  \end{split}
\end{equation}
Here, $\psi_{N\mathbf{k}}(\mathbf{r})$ and $E_{n\mathbf{k}}$ are quasiparticle wave function and quasiparticle energy with a band index $N$ and crystal momentum vector $\mathbf{k}$, respectively. 

Using $P_{QP}^{high}=P_{QP}-P_{QP}^{low}$ where $P_{QP}$ is RPA polarizability, the partially-screened Coulomb interaction ($W_r$) is calculated by
\begin{equation}
  \begin{split}
    W_r^{-1}(\mathbf{r},\mathbf{r}',\mathbf{k},i\nu_n)=V^{-1}(\mathbf{r},\mathbf{r}',\mathbf{k})-P_{QP}^{high}(\mathbf{r},\mathbf{r}',\mathbf{k},i\nu_n).
    \label{eq:partial_coulomb} 
  \end{split}
\end{equation}
Next, Slater's integrals ($F^k$) \cite{slater_quantum_1960,sugano_multiplets_2012} are calculated using $W_r(\mathbf{r},\mathbf{r}',\mathbf{R}=0,i\nu_n)$ and Wannier functions for the correlated orbitals. Here we note that Slater parameterization of Coulomb interaction tensor is an approximation by assuming full rotation symmetry.
\begin{equation}
  \begin{split}
    F^k(i\nu_n)=&\frac{1}{C_{k}^l}\frac{4\pi}{2k+1}\sum_{m_1,m_2,m_3,m_4}\sum_{m'_1,m'_2,m'_3,m'_4}\\
    &\langle Y_l^{m_1}|Y_k^{m_1-m_4}Y_l^{m_4}\rangle\langle Y_l^{m_2}Y_k^{m_2-m_3}|Y_l^{m_3}\rangle\\
    &S_{m_1m'_1}S_{m_2m'_2}S^{-1}_{m_3m'_3}S^{-1}_{m_4m'_4}
    \int d\mathbf{r}d\mathbf{r}'W_r(\mathbf{r},\mathbf{r}', R=0,i\nu_n)\\
    &W_{R=0,m'_1}^*(\mathbf{r})W_{R=0,m'_2}^*(\mathbf{r}')W_{R=0,m'_3}(\mathbf{r}')W_{R=0,m'_4}(\mathbf{r}),
    \label{eq:slater_woso}  
  \end{split}
\end{equation}
where $C_k^l=\frac{(2l+1)^4}{{2k+1}}\begin{bmatrix} l & k & l \\ 0 & 0 & 0\end{bmatrix}$, $\begin{bmatrix} l & k & l \\ 0 & 0 & 0\end{bmatrix}$ is the Racah-Wigner 3j-symbol, $S_{m_1m'_1}=\langle Y_{lm_1}|Y_{l}^{m}\rangle$, $|Y_{l}^{m}\rangle$ is spherical harmonics, and $|Y_{lm}\rangle$ is cubic spherical harmonics. $W_{\mathbf{R},m}(\mathbf{r})$ is a Wannier function for a correlated orbital with angular part of cubic spherical harmonics $Y_{lm}$ in the unicell at $\mathbf{R}$.


\section{Double counting energy}
The electron self-energy included in both \textit{ab initio} LQSGW and DMFT is the local Hartree term and the local GW term. They can be calculated as follows.
\begin{equation}
  \widetilde{\Sigma}_{DC,i,j}(i\omega_n)=2\sum_{k,l=m_l'}^{d\text{-orbital}} \widetilde{G}_{l,k}(\tau=0^-)\widetilde{\mathcal{U}}_{iklj}(i\nu=0)-\sum_{k,l}^{d\text{-orbital}}\int d\tau \widetilde{G}_{l,k}(\tau)\widetilde{W}_{ikjl}(\tau)e^{i \omega_n\tau}.\label{eq:dc}
\end{equation}
where $i$,$j$,$k$ and $l$ are orbital indices. $\widetilde{G}$ is the local Green's function. $\widetilde{\mathcal{U}}$ is constructed by using Slater's integrals in eq. \eqref{eq:slater_woso}. 
\begin{equation}
  \begin{split}
    \widetilde{\mathcal{U}}_{i,j,k,l}(i\nu_n)&=\sum_{\substack{m'_1m'_2,m'_3m'_4}}S_{i,m_1}S_{j,m_2}S_{k,m_3}^{-1}S_{l,m_4}^{-1}\\
    &\sum_{k=0}^{2l,even}\frac{4\pi}{2k+1}\langle Y_{l}^{m_1'}|Y_{k}^{q}Y_{l}^{m_4'}\rangle\langle Y_{l}^{m_2'}Y_{k}^{q}|Y_{l}^{m_3'}\rangle F^{k}(i\nu_n). 
    \label{eq:coulomb_so}
  \end{split}
\end{equation}

Here, we assume that the frequency-dependent interaction is of the form
\begin{equation}
  \widetilde{\mathcal{U}}_{ijkl}(i\nu_n) = \widetilde{U}_{ijkl} + F^0(i\nu_n)\delta_{il}\delta_{jk},
\end{equation}
that is, only the dynamical screening of the Slater-Condon parameter $F^0$ is taken into account. The other Slater-Condon parameters, which define $\widetilde{U}_{ijkl}$, are frequency independent and approximated by their value at $\nu_n=\infty$. $\widetilde{W}$ is the local screened Coulomb interaction given by
\begin{equation}
  \widetilde{W}_{ikjl}(i\nu_n){=}\widetilde{\mathcal{U}}_{ikjl}(i\nu_n)+\sum_{mnpq}^{d\text{-orbital}}\allowbreak \widetilde{\mathcal{U}}_{imnl}(i\nu_n) \allowbreak \widetilde{P}_{mpqn}(i\nu_n)\allowbreak \widetilde{W}_{pkjq}(i\nu_n),\label{eq:ww}
\end{equation}
where $\widetilde{P}$ is the local polarizability and it is calculated as
\begin{equation}
  \widetilde{P}_{mpqn}(i\nu_n)\allowbreak=\int d\tau
  \widetilde{G}_{n,p}(\tau)\widetilde{G}_{q,m}(-\tau)\allowbreak e^{i\nu_n\tau}.\label{eq:pi_wso}
\end{equation}

\section{DMFT self-consistent equation}
At each iteration of the fermionic DMFT self-consistent loop, the fermionic Weiss-field is constructed in the following way.
\begin{equation}
  \begin{split}
    \widetilde{\mathcal{G}}=\left(\left(\frac{1}{N_\mathbf{k}}\sum_\mathbf{k} f_\mathbf{k}^\dagger G(\mathbf{k},i\omega_n)f_\mathbf{k}\right)^{-1}+\widetilde{\Sigma}_{imp}\right)^{-1}
    \label{eq:glat} 
  \end{split}
\end{equation}
Here $f_\mathbf{k}$ is the fermionic projection operator to correlation orbitals (five Ni-d orbitals) and given by $f_\mathbf{k}=\langle\mathbf{r}|W_{i\mathbf{k}}\rangle$ where $|W_{i\mathbf{k}}\rangle=\frac{1}{\sqrt{N_\mathbf{k}}}\sum_{\mathbf{R}}|W_{i\mathbf{R}}\rangle e^{i\mathbf{k}\cdot\mathbf{R}}$. $\widetilde{\Sigma}_{imp}$ is impurity self-energy from impurity solver.
Within \textit{ab initio} LQSGW+DMFT, lattice Green's function is calculated by embedding impurity self-energy into the LQSGW Green's function
\begin{equation}
  \begin{split}
    G^{-1}(\mathbf{k},i\omega_n)=i\omega_n-H_{QP}^{nl}(\mathbf{k})-f_\mathbf{k} \widetilde{\Sigma}_{imp}(i\omega_n) f_\mathbf{k}^\dagger,
    \label{eq:glat_inv}
  \end{split}
\end{equation}
where $H_{QP}^{nl}$ is non-local LQSGW Hamiltonian\cite{tomczak_qsgw_2015}, in which double-counting self-energy is compensated up to linear order in frequency. 

\begin{equation}
  \begin{split}
    H_{QP}^{nl}(\mathbf{k})=\sqrt{Z_{DC}^{-1}(\mathbf{k})} H_{QP}\sqrt{Z_{DC}^{-1}(\mathbf{k})}-f_\mathbf{k} \widetilde{\Sigma}_{DC}(\omega=0) f_\mathbf{k}^\dagger.
    \label{eq:h_qp_nl} 
  \end{split}
\end{equation}
Here, $H_{QP}$ is Wannier interpolated LQSGW Hamiltonian into $15\times 15\times 15$ $k$-grid. $Z_{DC}^{-1}(\mathbf{k})=1-f_\mathbf{k}\left({\partial{\widetilde{\Sigma}_{DC}(\omega=0)}}/{\partial{i\omega_n}}\right)f_\mathbf{k}^\dagger$. 

ComDMFT necessitates the solution of an impurity model action. In ComDMFT, hybridization-expansion continuous-time quantum Monte Carlo (CTQMC) is adopted. CTQMC is a stochastic approach to obtain numerically exact solutions of an impurity model. An impurity model consists of a small interacting system, the impurity, immersed in a bath of non-interacting electrons. The action of the impurity model relevant for GW+DMFT reads

\begin{equation}
  \begin{split}
    \label{equ:Action}
    S = &-\iint_0^\beta \sum_{ij} c^\dagger_i(\tau) \widetilde{\mathcal{G}}_{ij}^{-1}(\tau - \tau') c_j(\tau') d\tau d\tau' \\
    &\quad\quad+\frac{1}{2}\iint_0^\beta \sum_{ijkl} c^\dagger_i(\tau) c^\dagger_j(\tau') \widetilde{\mathcal{U}}_{ijkl}(\tau - \tau') c_k(\tau')c_l(\tau) d\tau d\tau', 
  \end{split}
\end{equation}
where $c^\dagger_i$ creates an electron in the generalized orbital $i$ (which includes both spin and orbital degrees of freedom), $\beta$ is the inverse temperature, $\widetilde{\mathcal{G}}_{ij}$ is the fermionic Weiss field in eq. \eqref{eq:glat} and $\widetilde{\mathcal{U}}_{ijkl}$ in eq. \eqref{eq:coulomb_so}. 

We assume that the frequency-dependent interaction is of the form
\begin{equation}
  \widetilde{\mathcal{U}}_{ijkl}(i\nu_n) = \widetilde{U}_{ijkl} + F^0(i\nu_n)\delta_{il}\delta_{jk},
\end{equation}
that is, only the dynamical screening of the Slater-Condon parameter $F^0$ is taken into account, for the simplicity in the numerical algorithm based on hybridization-expansion CTQMC. The other Slater-Condon parameters, which define $\widetilde{U}_{ijkl}$, are frequency independent and approximated by their value at $\nu_n=\infty$. 

\section{orbital-resolved spectral function of undoped LaNiO$_2$}
\begin{figure}[H]
  \centering
  \includegraphics[width=0.5\textwidth]{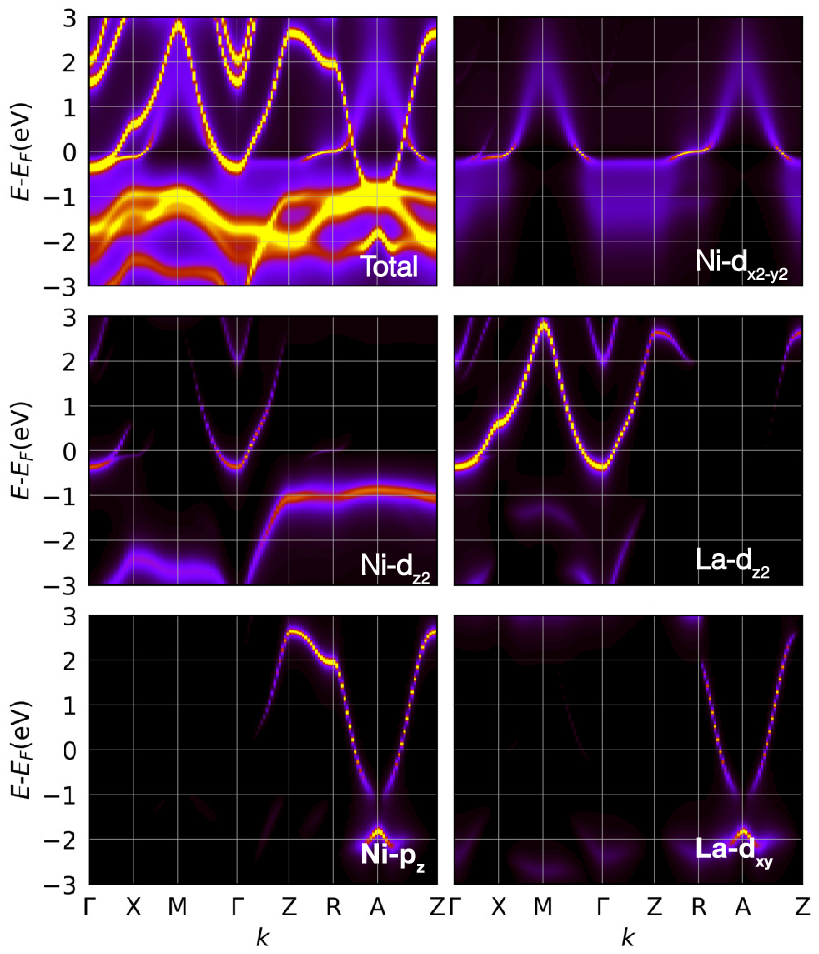}
  \caption{Total and orbital-resolved spectral function of LaNiO$_2$ along a high-symmetry line within \textit{ab initio} LQSGW+DMFT. Among the two bands crossing the Fermi level, the lower energy band shows Ni-$d_{x^2-y^2}$ character. The other self-doping band at higher energy is a mixture of four different orbitals. In the $\Gamma$-X-M-$\Gamma$ plane, its orbital characters are La-$d_{z^2}$ and Ni-$d_{z^2}$. In the Z-R-A-Z plane, its orbital characters are La-$d_{xy}$ and Ni-$p_{z}$.}
  \label{fig_spectral_undoped}
\end{figure}

Fig. \ref{fig_spectral_undoped} shows total and orbital-resolved spectral function of LaNiO$_2$ along a high-symmetry line within \textit{ab initio} LQSGW+DMFT. The bandstructure and orbital characters are essentially the same as those of La$_{0.8}$Ba$_{0.2}$NiO$_2$. the total spectral function shows that there are two bands crossing the Fermi level. Of these two bands, the lower energy band shows strong two-dimensional character, and it is dominated by the Ni-$d_{x^2-y^2}$ orbitals. The remaining band, the so-called self-doping band, is a higher energy band which shows strong hybridization between other Ni orbitals and La orbitals. The band dispersion varies strongly along the direction normal to the Ni-O plane ($\hat{z}$), demonstrating the strong 3-dimensional character of the self-doping band \cite{hepting_electronic_2020}. Moreover, the orbital character of the self-doping band is strongly dependent on $k_z$. In the $k_z$=0 plane, the orbital character of the self-doping band is mostly La-$d_{z^2}$ and Ni-$d_{z^2}$ \cite{lee_infinite-layer_2004,lechermann_late_2020}, In contrast, in the $k_z=\pi/c$ plane, where $c$ is the lattice constant along the $\hat{z}$ direction, its orbital character is mostly La-$d_{xy}$ and Ni-$p_{z}$.


\section{Fermi-surface}
\begin{figure}[h]
  \centering
  \includegraphics[width=1.0\textwidth]{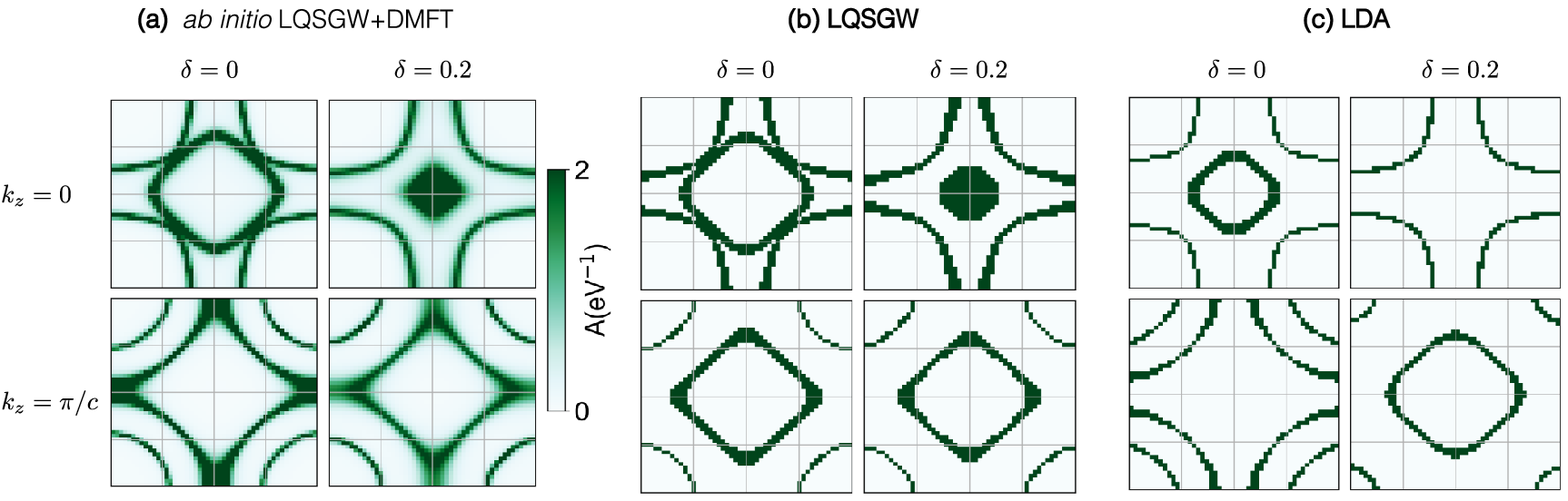}
  \caption{La$_{1-\delta}$Ba$_\delta$NiO$_2$ Fermi surface in the $k_z=0$ plane and $k_z=\pi/c$ plane within (a) \textit{ab initio} LQSGW+DMFT, (b) LQSGW and (c) LDA}
  \label{fig_fermi_surface}
\end{figure}

Fig. \ref{fig_fermi_surface} shows La$_{1-\delta}$Ba$_\delta$NiO$_2$ Fermi surface in the $k_z=0$ plane and $k_z=\pi/c$ plane and compares three different methodologies of \textit{ab initio} LQSGW+DMFT, LQSGW and LDA. In comparison to LDA, where the electron pocket in the $k_z=0$ disappears at the 0.2 Ba doping, the electron pocket within \textit{ab initio} LQSGW+DMFT and LQSGW is still present at the 0.2 Ba doping. Within \textit{ab initio} LQSGW+DMFT, two-dimensionality of Ni-$d_{x^2-y^2}$ hole-pocket is strong and it is robust against Ba-doping. As shown in Fig. \ref{fig_fermi_surface} (a), the position and size of Ni-$d_{x^2-y^2}$ hole-pocket in $k_z=0$ plane are similar to those in $k_z=\pi/c$ plane. In addition, these do not change by Ba-doping. This is in sharp contrast to LQSGW where two-dimensionality of the Ni-$d_{x^2-y^2}$ hole-pocket is weaker \cite{olevano_ab_2020} and LDA where Ni-$d_{x^2-y^2}$ hole-pocket shape is strongly affected by Ba doping.


\section{doping dependence of orbital occupation}

\begin{figure}[H]
  \centering
  \includegraphics[width=0.8\textwidth]{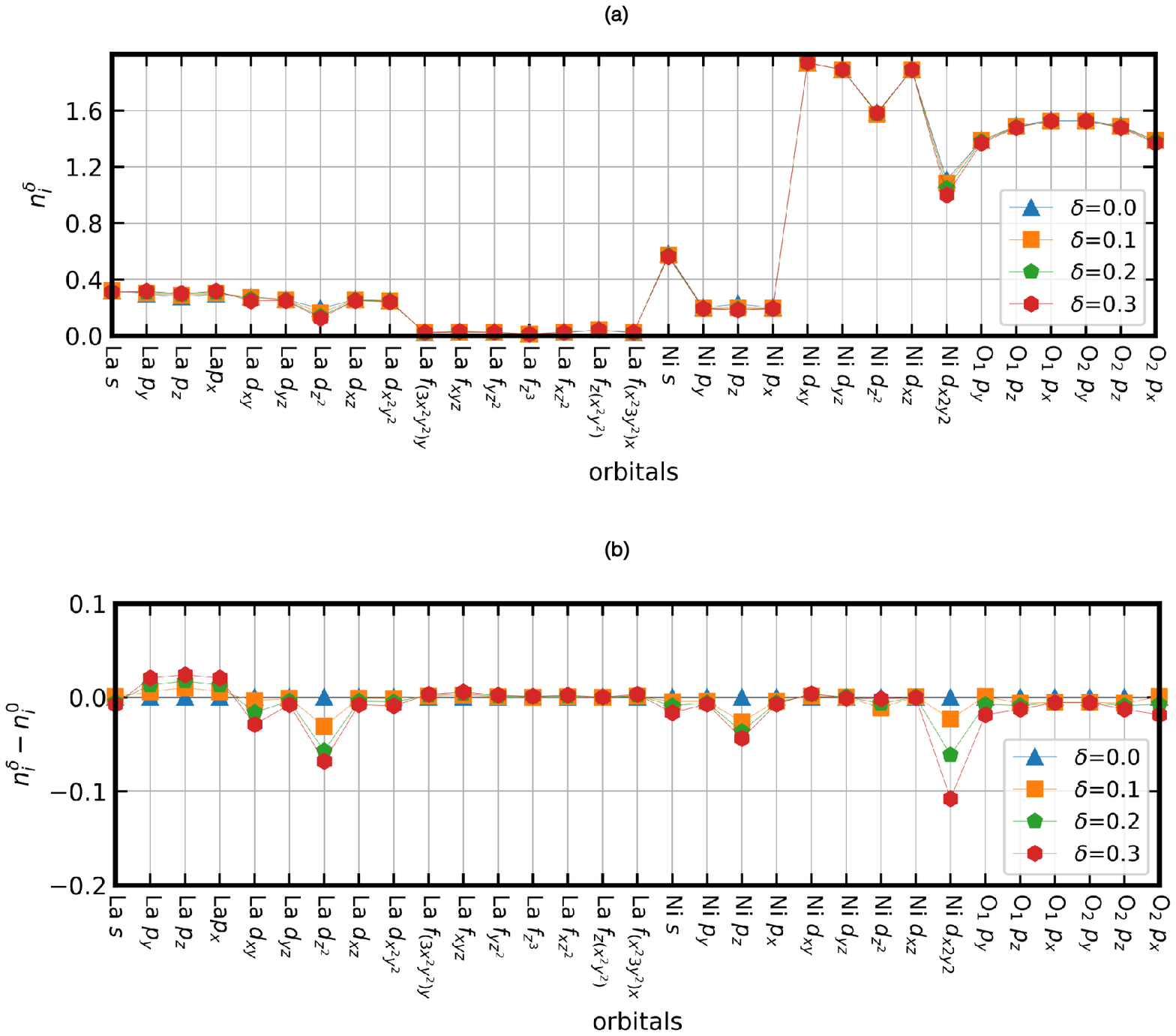}
  \caption{(a) The doping dependence of orbital occupation. (a) Doping induced orbital occupation change}
  \label{fig_occ}
\end{figure}

Fig. \ref{fig_occ} shows the doping dependence of orbital occupation. In the undoped compound, Ni-$t_{2g}$ orbitals are almost fully filled. In contrast, Ni-$e_{g}$ orbitals are partially filled. Due to the self-doping effect, Ni-$d_{x^2-y^2}$ orbital is away from half-filling. Upon Ba doping, the occupation of Ni-$d_{x^2-y^2}$ approaches to half-filling \cite{leonov_lifshitz_2020,petocchi_normal_2020}. Up to 30\% Ba doping, only $\sim$ 30\% of the doped hole goes into Ni-$d$ orbitals, especially Ni-$d_{x^2-y^2}$, but all the remaining holes go into other orbitals, especially La-$d_z$, La-$d_{xy}$ and Ni-$p_z$. 

\section{Other Ni-$e_g$ Hund's metal signitures in local susceptibilities}



\begin{figure}[H]
  \centering
  \includegraphics[width=0.6\textwidth]{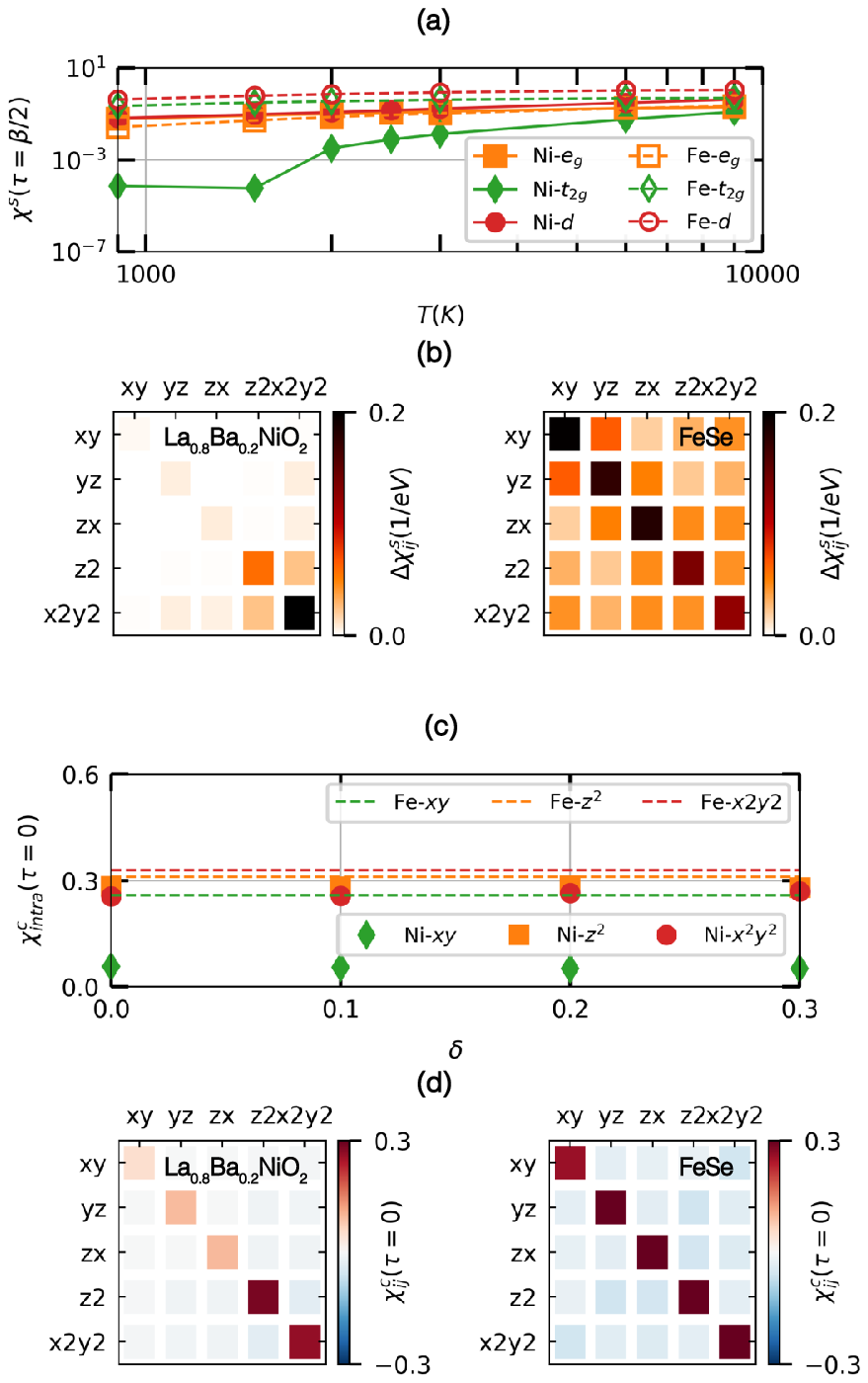}
  \caption{The temperature and doping dependence of the local spectrum of the charge, spin and orbital susceptibilities. (a) The temperature dependence of $\chi^s(\tau=\beta/2)$ of $d$ orbitals (red dots), $t_{2g}$ orbitals (green diamonds), and $e_g$-orbitals (orange squares) in La$_{0.8}$Ba$_{0.2}$NiO$_2$ (filled markers) and FeSe (empty markers). (b) Orbital-resolved dynamical contribution of spin susceptibility ($\Delta \chi_{ij}^s$) of Ni-$d$ orbitals in La$_{0.8}$Ba$_{0.2}$NiO$_2$ and Fe-$d$ orbitals in FeSe at T=900 K. (c) The doping dependence of instantaneous intraorbital charge susceptibility ($\chi_{ii}^s(\tau=0)$) of $d$ orbitals in La$_{0.8}$Ba$_{0.2}$NiO$_2$ (markers) and FeSe (dashed lines). (d) Instantaneous charge susceptibility ($\chi_{ij}^s(\tau=0)$) of Ni-$d$ orbitals in La$_{0.8}$Ba$_{0.2}$NiO$_2$ and Fe-$d$ orbitals in FeSe at T = 900 K.}
  \label{fig_sus_others}
\end{figure}

Five Ni-$d$ orbitals in La$_{1\mhyphen\delta}$Ba$_{\delta}$NiO$_2$ show enhanced dynamical spin susceptibility. According to Werner et al. \cite{werner_spin_2008}, at the spin-freezing crossover regime where Hundness dominates the electron correlation, the system shows enhanced $\chi_{tot}^s(\tau=\beta/2)=\sum_{ij=d}\langle S_{iz}(\beta/2)S_{jz}(0)\rangle$ \cite{leonov_lifshitz_2020} and it increases linearly upon heating. The coefficent of linear term can be related to NMR relaxation rates \cite{randeria_pairing_1992}. If Hundness gets stronger, the system shows sublinear temperature dependence. As shown in Fig \ref{fig_sus_others}(a), both Fe-$d$ (empty red dots) orbitals in FeSe and Ni-$d$ orbitals (filled red dots) in La$_{1\mhyphen\delta}$Ba$_{\delta}$NiO$_2$ show sublinear dependence in temperature. 

Five Ni-$d$ orbitals in La$_{1\mhyphen\delta}$Ba$_{\delta}$NiO$_2$ show orbital decoupling behavior, much like the Fe-$d$ orbitals in FeSe. According to de' Medici \cite{de_medici_hunds_2011}, a Hund's metal shows orbital decoupling and this manifests in the suppressed instantaneous interorbital charge susceptibility ($\chi_{i\ne j}^c(\tau=0)$) while the intraorbital susceptibility ($\chi_{i=j}^c(\tau=0)$) is finite \cite{isidori_charge_2019,fanfarillo_orbital-dependent_2016}. Here, $\chi_{ij}^c(\tau=0)=\lim_{\tau\to 0}\langle N_i(\tau)N_j(0)\rangle$. Fe-$d$ orbitals in FeSe show the predicted orbital decoupling, as shown in Fig. \ref{fig_sus_others} (d). $\chi_{i\ne j}^c(\tau=0)$ between any two pairs of Fe-$d$ orbitals is negative and strongly suppressed in comparison to any intraorbital one. Similarly, the interorbital, instantaneous charge susceptibility is suppressed in the Ni-$d$ orbitals of La$_{1\mhyphen\delta}$Ba$_{\delta}$NiO$_2$.

However, there is an important distinction between the Ni-$d$ orbitals in La$_{1\mhyphen\delta}$Ba$_{\delta}$NiO$_2$ and Fe-$d$ orbtials in FeSe: The $t_{2g}$ orbitals in La$_{1\mhyphen\delta}$Ba$_{\delta}$NiO$_2$ are inactive. The idea of inactive $t_{2g}$ orbitals in La$_{1\mhyphen\delta}$Ba$_{\delta}$NiO$_2$ can be demonstrated further by examining the local susceptibilities in the spin, orbital, and charge channels. This makes it difficult to conclude that La$_{1\mhyphen\delta}$Ba$_{\delta}$NiO$_2$ is an canonical $d$-orbital Hund's metal. Let us discuss.

The temperature dependence of $\chi^s(\tau=\beta/2)$ supports that the $t_{2g}$ orbitals are inactive, as shown in Fig. \ref{fig_sus_others} (a), which depicts $\chi_{t_2g}^s(\tau=\beta/2)=\sum_{ij=t_{2g}} \langle S_{iz}(\tau=\beta/2)S_{jz}(0)\rangle$. In contrast to the Fe-$d$ orbitals in FeSe where $\chi_{t_2g}^s(\tau=\beta/2)$ (empty green diamonds) shows sublinear temperature dependence similar to $\chi_{tot}^s(\tau=\beta/2)$, $\chi_{t_2g}^s(\tau=\beta/2)$ of the Ni-$d$ orbitals in La$_{0.8}$Ba$_{0.2}$NiO$_2$ (filled green diamonds) is essentially zero over a wide range of temperatures.

The dynamical contribution to the local spin susceptibility ($\Delta \chi_{ij}^s$) provides additional evidence that the Ni-$t_{2g}$ orbitals  are inactive. Here $\Delta \chi_{ij}^s=\int_0^\beta d\tau \{\chi_{ij}^s(\tau)-\chi_{ij}^s(\tau=\beta/2)\}$. According to Werner et al. \cite{werner_spin_2008}, in the spin-freezing crossover regime where Hundness dominates, the system has a large spin susceptibility ($\chi^s$) with a suppressed $\Delta \chi^s$. As shown in Fig. \ref{fig_sus_others} (b), all possible pairs of $\Delta \chi_{i,j}^s$ in FeSe are active. In contrast, the Ni-$e_g$ orbital contributions dominate in La$_{0.8}$Ba$_{0.2}$NiO$_2$.

Third, the instantaneous charge susceptibility exhibits intraorbital charge fluctuations primarily in the Ni-e$_{g}$ subspace. Fig. \ref{fig_sus_others} (c) and (d) show $\chi_{ii}^c(\tau=0)$. In FeSe, all five intraorbital components are similar in magnitude. In contrast, $\chi_{ii=Ni\text{-}t_{2g}}^c(\tau=0)$ is strongly suppressed in comparison to $\chi_{ii=Ni\text{-}e_g}^c(\tau=0)$ in La$_{0.8}$Ba$_{0.2}$NiO$_2$.

Once we narrow down our view from all Ni-$d$ orbitals into only the Ni-$e_g$ orbitals, we can successfully find all signatures of a Hund's metal in the temperature and doping dependence of the local spectrum of the charge, spin and orbital susceptibilities. Ni-$e_g$ orbitals in La$_{1\mhyphen\delta}$Ba$_{\delta}$NiO$_2$ show enhanced dynamical spin susceptibility. $\chi_{e_g}^s(\tau=\beta/2)=\sum_{ij=e_{g}} \chi_{ij}^s(\tau=\beta/2)$ in Fig. \ref{fig_sus_others}(a) demonstrates that $\chi_{e_g}^s(\tau=\beta/2)$ in Ni atom (filled orange squares) has a sublinear temperature dependence similar to $\chi_{tot}^s(\tau=\beta/2)$. The Ni-$e_g$ orbital contributions dominate $\Delta \chi_{i,j}^s$ as shown in Fig. \ref{fig_sus_others} (b). Ni-$e_g$ orbitals in La$_{1\mhyphen\delta}$Ba$_{\delta}$NiO$_2$ show orbital decoupling behavior as the Fe-$d$ orbitals in FeSe. As shown in Fig. \ref{fig_sus_others}(d), $\chi_{x^2-y^2,z^2}^c(\tau=0)$ is strongly suppressed. Thus we conclude  that La$_{1\mhyphen\delta}$Ba$_{\delta}$NiO$_2$ is a $e_g$ Hund's metal: That is, the Hundness exists only among the Ni-$e_g$ orbitals.

\section{Hubbard band Energy gap in the two-orbital Kanamori Hamiltonian}
\begin{table}[t]%
  \centering
  \caption{The eigenstates and eigenenergies of two-orbital Kanamori Hamiltonian in its atomic limit. Here, $a=J/\left(\left(\sqrt{J^2+\Delta^2}-\Delta\right)^2+J^2\right)$ and $b=\sqrt{1-a^2}$}
  \begin{tabularx}{0.8\columnwidth}{cYYYc}

    \hline
    \hline 
    State     & $N_{e_g}$ & $S_{e_g}$ & $S_{e_g}^z$ & Energy \\
    \hline
    \hline 
    $|0, 0\rangle$     & 0 & 0 & 0 & 0\\
    \hline  
    $|0, \uparrow\rangle$     & 1 & 1/2 & 1/2 & $-\Delta$\\
    $|0, \downarrow\rangle$     & 1 & 1/2 & -1/2 & $-\Delta$\\
    $|\uparrow, 0\rangle$     & 1 & 1/2 & 1/2 & 0\\
    $|\downarrow, 0\rangle$     & 1 & 1/2 & -1/2 & 0\\
    \hline
    $|\uparrow, \uparrow\rangle$     & 2 & 1 & 1 & $-\Delta+U'-J$\\ 
    $\left(|\uparrow, \downarrow\rangle+|\downarrow, \uparrow\rangle\right)/\sqrt{2}$     & 2 & 1 & 0 & $-\Delta+U'-J$\\
    $|\downarrow, \downarrow\rangle$     & 2 & 1 & -1 & $-\Delta+U'-J$\\
    $\left(|\uparrow, \downarrow\rangle-|\downarrow, \uparrow\rangle\right)/\sqrt{2}$     & 2 & 0 & 0 & $-\Delta+U'+J$\\
    $\left(b|\uparrow\downarrow,0\rangle-a|0, \uparrow\downarrow\rangle\right)$     & 2 & 0 & 0 & $-\Delta+U-\sqrt{J^2+\Delta^2}$\\   
    $\left(a|\uparrow\downarrow,0\rangle+b|0, \uparrow\downarrow\rangle\right)$     & 2 & 0 & 0 & $-\Delta+U+\sqrt{J^2+\Delta^2}$\\
    \hline
    $|\uparrow, \uparrow\downarrow\rangle$     & 3 & 1/2 & 1/2 & $-2\Delta+U+2U'-J$\\
    $|\downarrow, \uparrow\downarrow\rangle$     & 3 & 1/2 & -1/2 & $-2\Delta+U+2U'-J$\\
    $|\uparrow\downarrow, \uparrow\rangle$     & 3 & 1/2 & 1/2 & $-\Delta+U+2U'-J$\\
    $|\uparrow\downarrow, \downarrow\rangle$     & 3 & 1/2 & -1/2 & $-\Delta+U+2U'-J$\\
    \hline
    $|\uparrow\downarrow, \uparrow\downarrow\rangle$     & 4 & 0 & 0 & $-2\Delta+2U+4U'-2J$\\
    \hline 
    \hline 
  \end{tabularx}
  \label{tab_kanamori}
\end{table}

Two-orbital Kanamori Hamiltonian with vanishing intersite hopping is given by

\begin{equation}
  \begin{split}
    H &= -\Delta \sum_\sigma n_{2\sigma}+U\sum_{i}{n_{i \uparrow} n_{i \downarrow}} 
    + \sum_{i,j,\sigma,\sigma'}^{i \neq j}\left(U'-J\delta_{\sigma,\sigma'}\right) n_{ i \sigma} n_{ j \sigma'} 
    - J\sum_{i,j}^{i \neq j}\left(c_{i \uparrow}^{\dagger} c_{i \downarrow} c^{\dagger}_{j \downarrow} c_{j \uparrow} 
      - c^{\dagger}_{ i \uparrow} c^{\dagger}_{ i \downarrow} c_{ j \downarrow} c_{ j \uparrow}\right) \\, 
    \label{eq_kanamori}
  \end{split}
\end{equation}
Here, $\Delta$, $U$, $U'$ and $J$ are the crystal-field splitting which is positive, intraorbital Coulomb interaction, interorbital Coulomb interaction, and Hund's coupling, respectively. The eigenenergies and eigenstates of the Hamiltonian is tabulated in Table \ref{tab_kanamori}.

The separation between upper and lower Hubbard bands ($U_{eff}$) at a fractional occupation between $2\leq N_{e_g} \leq 3$ might be obtained by the ensemble average of the separation.
\begin{equation}
  \begin{split}
    U^{eff}(2 \leq N_{e_g} \leq 3)&=U^{eff}(N_{e_g}=2)(3-N_{e_g})+U^{eff}(N_{e_g}=3)(2-N_{e_g})\\
    \label{gap_fractional}  
  \end{split}
\end{equation}
If $\Delta >\sqrt{(U-U'+J)^2-J^2}$, the singlet state with $E=-\Delta+U-\sqrt{J^2+\Delta^2}$ is the ground state in $N_{e_g}$=2 subspace and the Hubbard band separation is given by

\begin{equation}
  \begin{split}
    U^{eff}(N_{e_g}=2)&=-J-U+2U'-\Delta+2\sqrt{J^2+\Delta^2}\\
    U^{eff}(N_{e_g}=3)&=U+\Delta-\sqrt{J^2+\Delta^2}\\
  \end{split}
  \label{gap_fractional}   
\end{equation}`

If $\Delta <\sqrt{(U-U'+J)^2-J^2}$, the triplets are the ground states in $N_{e_g}$=2 subspace and Hubbard band separation is given by
\begin{equation}
  \begin{split}
    U^{eff}(N_{e_g}=2)&=U+J-\Delta\\
    U^{eff}(N_{e_g}=3)&=U'-J+\Delta\\ 
  \end{split}
  \label{gap_fractional}  
\end{equation}`

Then the Hubbard band separation at a fractional occupation between $2 \leq N_{e_g} \leq 3$ is given by
\begin{equation}
  U^{eff}(2 \leq N_{e_g} \leq 3)=
  \begin{cases}
    \left(-U+U'+2\Delta-2J\right)N_{e_g}\\+3U-2U'+5J-5\Delta, &\text{if}\ \Delta <\sqrt{(U-U'+J)^2-J^2} \\
    \\
    \left(J+2U-2U'+2\Delta-3\sqrt{J^2+\Delta^2}\right)N_{e_g}\\-5U+6U'-5\Delta-3J+8\sqrt{J^2+\Delta^2}, &\text{otherwise} \\
  \end{cases}
  \label{gap_fractional}  
\end{equation}`

The Hubbard band separation at a fractional occupation between $2 \leq N_{e_g} \leq 3$  can be written as 
\begin{equation}
  U^{eff}=U_{\Delta=0}^{eff}+
  \begin{cases}
    (2N_{e_g}-5)\Delta, &\text{if}\ \Delta <\sqrt{(U-U'+J)^2-J^2} \\ 
    (2N_{e_g}-5)\Delta-(3N_{e_g}-8)(\sqrt{J^2+\Delta^2}-J), &\text{otherwise} \\
  \end{cases}
  \label{gap_fractional}  
\end{equation}
Here, $U_{\Delta=0}^{eff}$ denotes the Hubbad band gap when $\Delta=0$. If $2.5 \leq N_{e_g} \leq 3$, $U_{\Delta=0}^{eff} \leq U^{eff}$. 

\section{Virtual crystal approximation and its justification}
Virtual crystal approximation is a tractable and cost-efficient way of studying configurationally disordered systems. Instead of performing an ensemble average over various possible atomic arrangement in large supercells, average effective potential in a crystal with a primitive periodicity is employed to calculate physical properties. Virtual crystal approximation has been widely used in various correlated electron systems. Especially, its usage in the Fe-based superconductor, which is a prototypical Hund's metal system, has been justified by comparing various physical quantities including dynamical magnetic susceptibility \cite{li_orbital_2016,xu_strong_2020}.

DFT+sicDMFT calculations of Nd$_{1-\delta}$Sr$_{\delta}$NiO$_2$ in 2$\times$2$\times$2 supercells \cite{lechermann_late_2020} reported several results consistent with ours based on virtual crystal approximation. First, doping-induced structural relaxation turns out to induce only minor modulation. Sr-doping leads to a minor modulation of mainly the Ni-O bond lengths, up to only 0.5 $\%$ of the respective stoichiometric distances. Second, electron occupation at $\delta=0.25$ within the supercell approach results in almost identical to that of our results at $\delta=0.2$. Within the supercell approach, electron occupation in Ni-$d_{x^2-y^2}$ and Ni-$d_{z^2}$ are 1.09 and 1.61, respectively. These values are almost identical to ours of 1.04 and 1.60. Third, Ni-$t_{2g}$ orbital is almost fully filled in both calculations.

In addition, virtual crystal approximation is expected to be valid in the low-doping concentration limit, where the effect of substitutional doping is minor. If we compare our spectral function in the doped system with one in the undoped system, their electronic structures are essentially identical, justifying our usage of virtual crystal approximation.

\section{Ni-$p_z$ orbital in the low-energy electronic structure}
\begin{figure}[H]
  \centering
  \includegraphics[width=0.8\columnwidth]{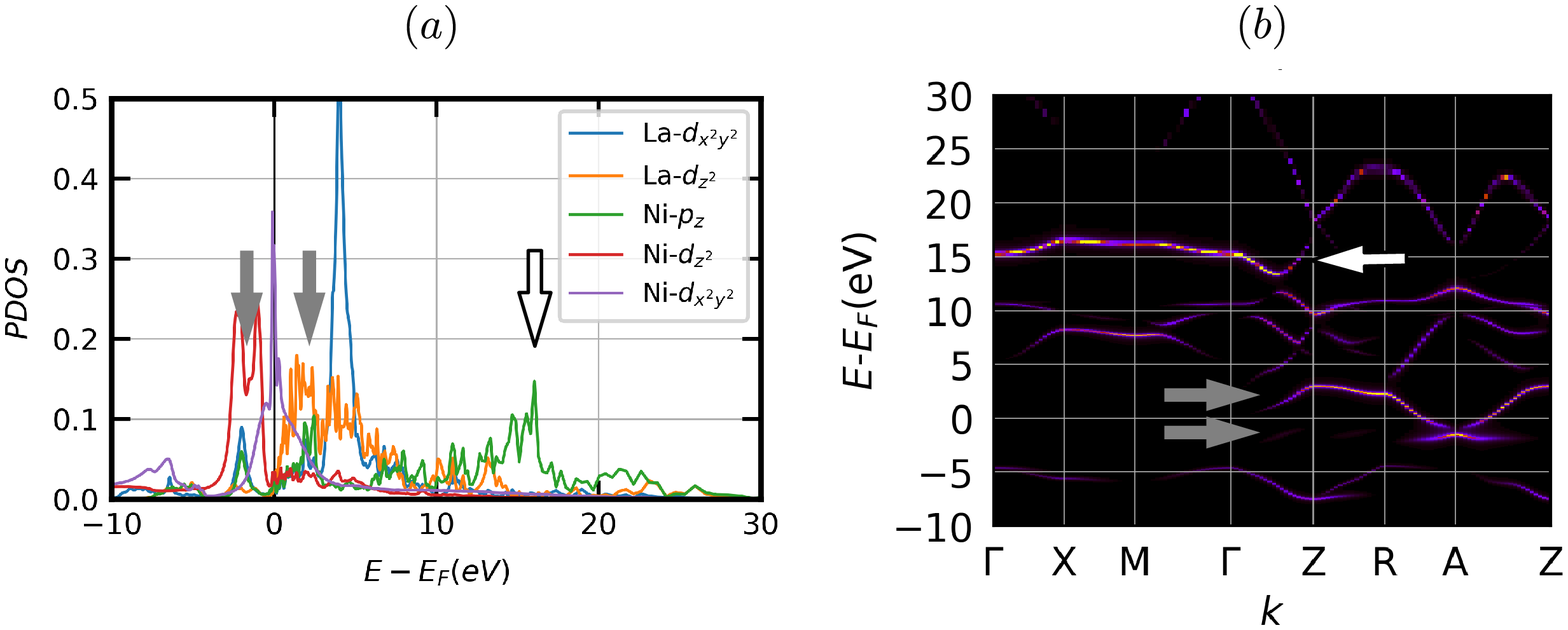}
  \caption{(a) Projected density of states (PDOS) to the La-$d_{xy}$, La-$d_{z^2}$, Ni-$p_z$, Ni-$d_{z^2}$, and Ni-$d_{x^2-y^2}$ within \textit{ab initio} LQSGW+DMFT at T=300K. The locations of the Ni-$p_z$ main peak as well as two subpeaks are marked by white and grey arrows. (b) Ni-$p_{z}$ orbital projected spectral function of La$_{0.8}$Ba$_{0.2}$NiO$_2$ along a high-symmetry line. The relatively flat bands contributing to the Ni-$p_{z}$ peaks in the PDOS are marked by arrows with the same colors. }
  \label{fig_pz}
\end{figure}

The inclusion of Ni-$p_z$ orbital in the basis set and its substantial contribution to the low-energy physics, especially in the Z-R-A-Z plane, is originated from our choice of relatively large frozen energy window ($-10eV <E-E_F<11eV $) during the Wannier function constructions. Green lines in Fig. \ref{fig_pz} (a) shows Ni-$p_z$ projected density of states. Its main peak marked by the white arrow is located at $E-E_F=16eV$. Due to its relatively high energy, this orbital is often neglected in the basis set. However, there is two subpeak near the Fermi levels (marked by grey arrows), which are non-negligible. As shown in the Ni-$p_z$ orbital-resolved spectral function in \ref{fig_pz} (b), there are relatively flat bands at each peak energy.

Without Ni-$p_z$ orbital in the basis set, we can still reproduce low-energy bandstructure. However, the low-energy bandstructure will match to the \textit{ab initio} bandstructure in a narrower energy window than $-10eV <E-E_F<11eV $. In addition, the size of the other orbitals than Ni-$p_z$ orbital, such as Ni-$d$ orbitals and La-$d$ orbitals, will be larger than the size of the same orbitals constructed together with Ni-$p_z$ orbital.

\section{cRPA schemes and Ni-$e_g$ Hund's physics}
\begin{figure}[H]
  \centering
  \includegraphics[width=0.8\columnwidth]{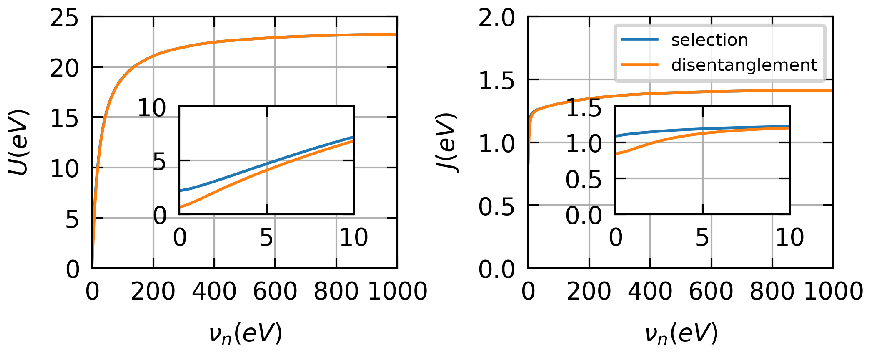}
  \caption{(a) $U$ and (b) $J$ obtained from two different constrained random phase approximation schemes: band selection method \cite{sakuma_first-principles_2013,amadon_screened_2014, werner_satellites_2012} and disentanglement method \cite{miyake_ab_2009}.}
  \label{fig_wnds_uj}
\end{figure}

Within LQSGW+DMFT approach, the Coulomb interaction tensor is calculated within cRPA. Within this method, the choice of $P_{QP}^{low}$ is not unique and there are several different schemes proposed. We tested the robustness of our results by employing another cRPA scheme: disentanglement method\cite{miyake_ab_2009}. 

\begin{figure}[H]
  \centering
  \includegraphics[width=0.48\columnwidth]{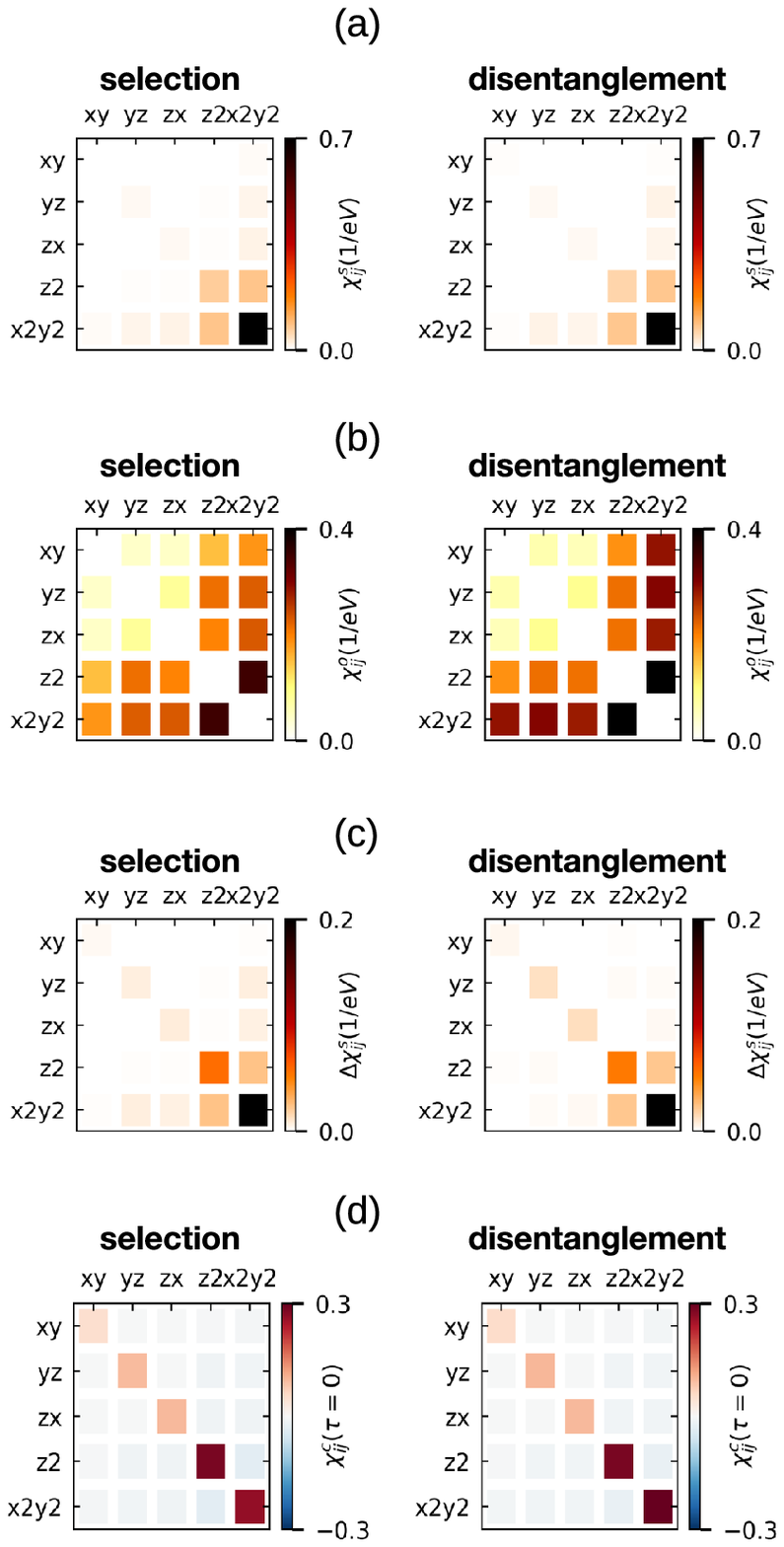}
  \caption{(a) Orbital-resolved static spin susceptibility ($\chi_{ij}^s$), (b) orbital susceptibility ($\chi_{ij}^s$), (c) orbital-resolved dynamical contribution of spin susceptibility ($\Delta \chi_{ij}^s$) and (e) orbital-resolved instantaneous charge susceptibility ($\chi_{ij}^s(\tau=0)$) obtained with two different constrained random phase approximation schemes: band selection method \cite{sakuma_first-principles_2013,amadon_screened_2014, werner_satellites_2012} and disentanglement method \cite{miyake_ab_2009}.}
  \label{fig_wnds}
\end{figure}

As shown in Fig. \ref{fig_wnds_uj}, the band selection method \cite{sakuma_first-principles_2013,amadon_screened_2014, werner_satellites_2012} employed in our paper provides slightly larger $U$ and $J$ values in its static value than those from the disentanglement method. However, $J$ values and their frequency dependence is relatively insensitive to the definition of choice of $P_{low}^{QP}$ in comparison with $U$ values. By using this $U$ and $J$ obtained from the disentanglement method, we recalculated local spin and orbital susceptiblity and compared the results with the one from the band selection method. As shown in Fig. \ref{fig_wnds}, we were not able to find any meaningful changes in the local susceptiblity in spin orbital and charge degree of freedoms, which are important measures to confirm Hund's metallicity.

\section{The position of the self-doping band at the $\Gamma$ point and Hund's metal physics.}
\begin{figure}[h]
  \centering
  \includegraphics[width=0.6\textwidth]{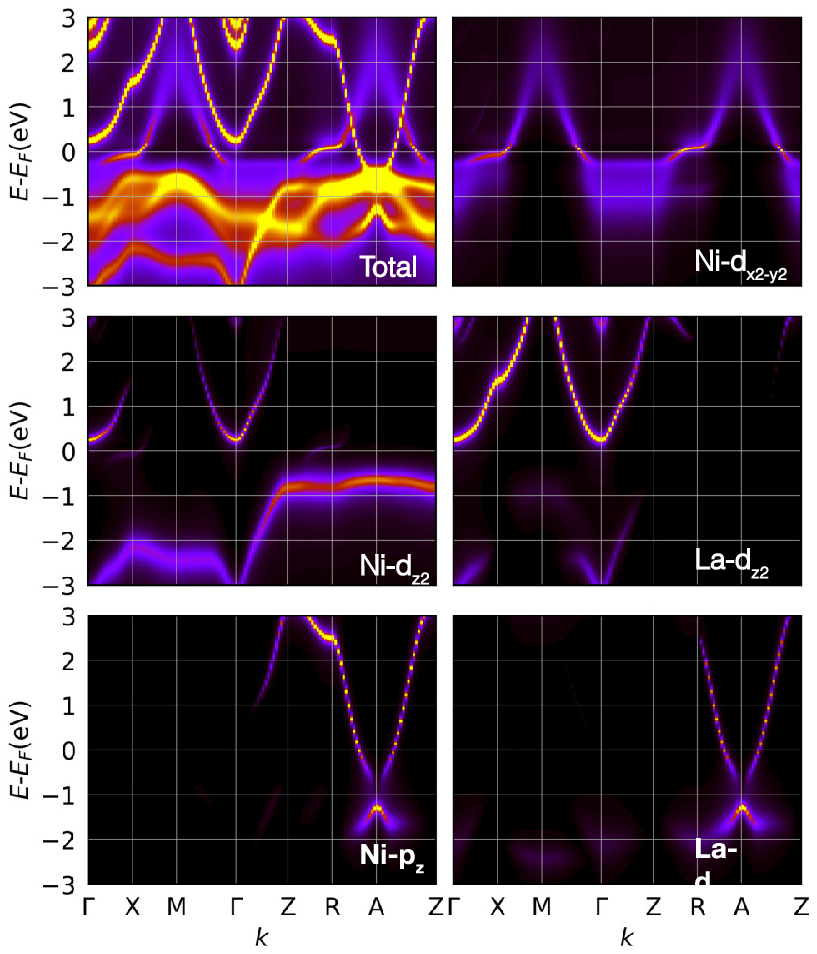}
  \caption{Total and orbital-resolved spectral function of La$_{0.7}$Ba$_{0.3}$NiO$_2$ along a high-symmetry line as calculated within \textit{ab initio} LQSGW+DMFT at T=300K. Of the two bands crossing the Fermi level, the lower energy band shows Ni-$d_{x^2-y^2}$ character, and the other, self-doping band at higher energy is a mixture of La-$d_{z^2}$, Ni-$d_{z^2}$, La-$d_{xy}$ and Ni-$p_{z}$. } 
  \label{fig_spectral03}
\end{figure} 

Depending on the choice of the \textit{ab initio} method, the location of the self-doping band at the $\Gamma$ point of the undoped infinite-layer nickelates varies. This may change Ni-$d$ electron occupation and play a role in the proposed Ni-$e_g$ Hund physics. There is a recent DFT+DMFT report \cite{si_topotactic_2020} that the inclusion of the electron correlation in La-$d$ orbital within dynamical mean-field theory shifts the self-doping band above the Fermi level. Although orbital occupation is not stated in the paper, the PDOS plot in Fig. S.3(b) of the paper indicates that Ni-$d_{z^2}$ is almost fully occupied within their approach, implying $d^9$ occupation in the Ni-d orbitals, in support of the Mott-Hubbard picture.

However, this result is in sharp contrast to the results within our LQSGW+DMFT calculation on LaNiO$_2$ as well as multitier-GW+EDMFT calculation on NdNiO$_2$ \cite{petocchi_normal_2020}. Here we note that La-$d$ is treated within GW approximation in the LQSGW+DMFT calculation and within extended dynamical mean field theory in the multitier-GW+EDMFT calculation. In both parameter-free calculations, the self-doping band is below the Fermi level at $\Gamma$ point in the undoped compound.

A similar shift of the self-doping bands at the $\Gamma$ point above the Fermi level has been induced by Sr doping in both LQSGW+DMFT and multitier-GW+EDMFT calculations. To illustrate, the self-doping band is above the Fermi level at 30$\%$ Sr doping within the LQSGW+DMFT approach as shown in Fig. \ref{fig_spectral03}. At 20$\%$ Sr doping, the self-doping bands is above the Fermi level at the $\Gamma$ point within the multitier GW+EDMFT approach. More importantly, at these doping levels, Ni-d orbital is still far from $d^9$ occupation. To illustrate, Ni-d occupation is 8.31 at 30$\%$ Sr doping within multitier-GW+EDMFT and 8.32 at 20$\%$ Sr doping within LQSGW+DMFT approach. This result is in support of the Hund's metal picture.

\bibliography{zotero}